\documentclass[modern]{aastex631}

\newcommand{\ecss}{erg~cm$^{-2}$~s$^{-1}$~sr$^{-1}$}
\newcommand{\ecssa}{erg~cm$^{-2}$~s$^{-1}$~sr$^{-1}$~\AA$^{-1}$}
\newcommand{\pcssa}{photon~cm$^{-2}$~s$^{-1}$~sr$^{-1}$~\AA$^{-1}$}

\newcommand{\as}{$^{\prime\prime}$}
\renewcommand{\ion}[2]{#1\,\textsc{#2}}
\newcommand{\hinode}{Hinode}

\submitjournal{ApJ}

\graphicspath{{./}{figures/}}
\begin{document}

\title{Modeling Flare Continuum Emission Observed by Hinode/EIS: Instrument Calibration and Element Composition Results}

\author[0000-0001-9034-2925]{Peter R. Young}
\affiliation{NASA Goddard Space Flight Center, Heliophysics Division,
Greenbelt, MD 20771, USA}
\affiliation{Department of Mathematics, Physics and Electrical Engineering, Northumbria University, Newcastle upon Tyne, NE1 8ST, UK}

\author[0000-0002-7020-2826]{Biswajit Mondal}
\affiliation{NASA Postdoctoral Program, NASA Marshall Space Flight Center, ST13, Huntsville, AL 35812, USA}
\affiliation{University of Alabama in Huntsville, Huntsville, AL 35805, USA}

\begin{abstract}
Continuum emission from a solar flare 
observed with the Extreme ultraviolet Imaging Spectrometer (EIS) on board the Hinode satellite is used to obtain the radiometric calibration of the instrument. The flare had a GOES class of M8, and peaked at 23:59~UT on 2024 September 30. The continuum is modeled by computing a differential emission measure curve using EIS emission lines and atomic data from the CHIANTI database. The ratio of the observed continuum to model continuum yields effective area curves for the instrument. The new curves confirm earlier findings that the EIS long-wavelength channel has degraded by a factor two compared to the short-wavelength channel. However, no evidence is found for the fine-scale structure in the effective area curves that has been presented by previous authors.
In order to reproduce both the emission line intensities and the continuum, it is found that the plasma must be depleted in elements with low first ionization potentials (FIPs), i.e., the so-called inverse FIP-effect. In particular, the Fe/H relative abundance is found to be a factor 0.57 below the photospheric value at a temperature of 10~MK. This is confirmed by analysis of soft X-ray spectra from the Solar X-ray Monitor on Chandrayaan-2, which yields an Fe/H FIP bias of 0.55 averaged over the entire flare.
\end{abstract}

%---------------------
\section{Introduction}

An accurate radiometric calibration is critical to deriving physical parameters from an astrophysical spectrum. The extreme ultraviolet (EUV) part of the solar spectrum is observed by a number of space-borne instruments due to the wide range of diagnostics of the solar atmosphere in this range. Atmospheric absorption of EUV radiation requires that instruments must be flown in space, but it is not feasible to fly a calibration lamp with EUV instruments. The typical approach for the radiometric calibration is to perform a pre-launch calibration in the laboratory and then perform indirect checks in orbit. 
In flight, the calibration can be checked against existing space instruments, or by occasionally flying instruments on sounding rockets. For example, \citet{2014ApJS..213...11W} obtained an absolute radiometric calibration for the EUV Imaging Spectrometer \citep[EIS:][]{2007SoPh..243...19C} on Hinode by comparing EIS scans of the entire solar disk in the \ion{Fe}{xii} 195.12~\AA\ emission line with irradiance measurements from the EUV Variability Experiment on board the Solar Dynamics Observatory (SDO) over the period 2010--2013. The EUNIS rocket experiment was launched in 2007 and \citet{2011ApJS..197...32W} performed a cross-calibration with EIS. The EVE sounding rocket instrument is launched regularly to help calibrate the EVE instrument on SDO, which in turn allows calibration of the AIA instrument on SDO \citep{2014SoPh..289.2377B}, and hence calibrate EIS \citep{2025ApJS..276...42D}. 
The spectra from EUV spectrometers can also be self-calibrated by making use of line ratios that are insensitive to plasma conditions \citep{1983A&A...128..181N, 1998A&A...329..291Y,2013A&A...555A..47D}, although this method only gives the relative calibration across the wavelength range.

EIS was launched on the Hinode spacecraft in 2006, and the data have been used in over 500 refereed science articles. EIS obtains spectra in two wavelength channels covering 171--212~\AA\ and 246--292~\AA. The pre-launch calibration was described in \citet{2006ApOpt..45.8689L}, and post-launch updates were described in \citet{2013SoPh..282..629M}, \citet{2013A&A...555A..47D}, \citet{2014ApJS..213...11W} and \citet[][hereafter DZWW25]{2025ApJS..276...42D}. \citet{2013SoPh..282..629M} modeled an exponential decay in the EIS sensitivity that was independent of wavelength, but this was superseded by the analyses of \citet{2013A&A...555A..47D} and  \citet{2014ApJS..213...11W} when it was clear that the long wavelength channel was losing sensitivity more quickly than the short wavelength channel. These two works applied different approaches to deriving the EIS sensitivity as a function of time. \citet{2013A&A...555A..47D} made use of insensitive line ratios, while \citet{2014ApJS..213...11W} performed differential emission measure (DEM) analyses on off-limb quiet Sun datasets. 

DZWW25 combined the methods of \citet{2013A&A...555A..47D} and  \citet{2014ApJS..213...11W} to yield an updated time-dependent calibration covering the period 2007 to 2022. The absolute calibration was obtained by normalizing against the 193~\AA\ channel of AIA.

The present work does not supersede that of DZWW25, but obtains new effective area curves that apply on a single date, 2024 September 30. The method is entirely different in that the EUV continuum measured from a solar flare constrains the effective area curves across both EIS channels. This method was suggested as a method for calibrating EIS by \citet{2005A&A...441.1211F}, but has not previously been applied. Since the continuum emission is fairly uniform over small wavelength ranges, 
it is particularly valuable for investigating the fine-scale structure of the DZWW25 effective area curves, such as the ``shoulders" seen in Figures~4 and 5 of this work.

The EUV continuum also enables the absolute element abundances of the flare plasma to be determined by modeling the continuum and emission lines together. This method is commonly applied at X-ray wavelengths below around 10\,\AA\ \citep[e.g.,][]{2003ApJ...589L.113P,2015ApJ...803...67D,2021ApJ...912L..12V}. It has also been used in the soft X-ray/EUV region by \citet{2014ApJ...786L...2W} and the far-UV region by \citet{2007ApJ...659..743L}. The present article is the first time this approach has been applied to EIS spectra.

The present article is structured as follows: the availability of the data and software used in the analysis are described in Section~\ref{sec:data}, and the properties of the EUV flare continuum are given in Section~\ref{sec:cont}. Information about the EIS dataset  and flare are given in Section~\ref{sec:dataset}, and the procedure for preparing the data for analysis is described in Section~\ref{sec:prep}. The analysis procedure for yielding the effective area curves is presented in Section~\ref{sec:proc} and results are given in Section~\ref{sec:results}. The EIS analysis yields an estimate of the element composition of the plasma which is compared with the value obtained from simultaneous X-ray spectra in Section~\ref{sec:xsm}. A discussion of the effective area results is provided in Section~\ref{sec:discussion} and a final summary is given in Section~\ref{sec:summary}.

%=======================================
\section{Data and Software Availability}\label{sec:data}

Data from EIS and the Atmospheric Imaging Assembly \citep[AIA:][]{2012SoPh..275...17L} on board the Solar Dynamics Observatory are used in this article and they are available at the Virtual Solar Observatory\footnote{\url{https://virtualsolar.org}}. Additional data come from  the Solar X-ray Monitor \citep[XSM:][]{2020SoPh..295..139M,Mithun_2021ExA....51...33M} on Chandrayaan-2 and are available at the XSM website\footnote{\url{https://www.prl.res.in/ch2xsm/}}, along with the  XSM Data Analysis Software \citep[XSMDAS:][]{Mithun_2021A&C....3400449M}.  The EIS data were processed from level-0 to level-1 using IDL software available in the Solarsoft distribution \citep{1998SoPh..182..497F,2012ascl.soft08013F}. IDL software for spatially binning EIS spectra and performing Gaussian fits are described in the following sections and are also available in Solarsoft. Atomic data and synthetic spectrum modeling software are from version 11.0.2 of the CHIANTI atomic database \citep{2024ApJ...974...71D}. The DEM  is derived using the Markov Chain Monte Carlo (MCMC) method \citep{1998ApJ...503..450K}, and the software is available through the IDL  PINTofALE software package\footnote{\url{https://hea-www.harvard.edu/PINTofALE/}}. The \textsf{ch\_dem} package \citep{peter_young_2025_15396769} is used to interface between the MCMC code, CHIANTI, and the measured EIS line intensities. 

Software developed for the data analysis performed in this work is available in a GitHub repository (DOI:\href{https://doi.org/10.5281/zenodo.19800920}{10.5281/zenodo.19800920}).  The IDL script \textsf{continuum\_process\_script} is provided that executes all of the steps to derive the new effective area curves from the input intensities. The \textsf{data} sub-directory contains data files, including the measured emission line and continuum intensities, DEM solutions, and final effective area solutions. The sub-directory \textsf{figures} contains the IDL routines for generating the figures used in this article.

%=============================
\section{Continuum properties}\label{sec:cont}

\begin{figure}[h]
    \centering
    \includegraphics[width=1.0\linewidth]{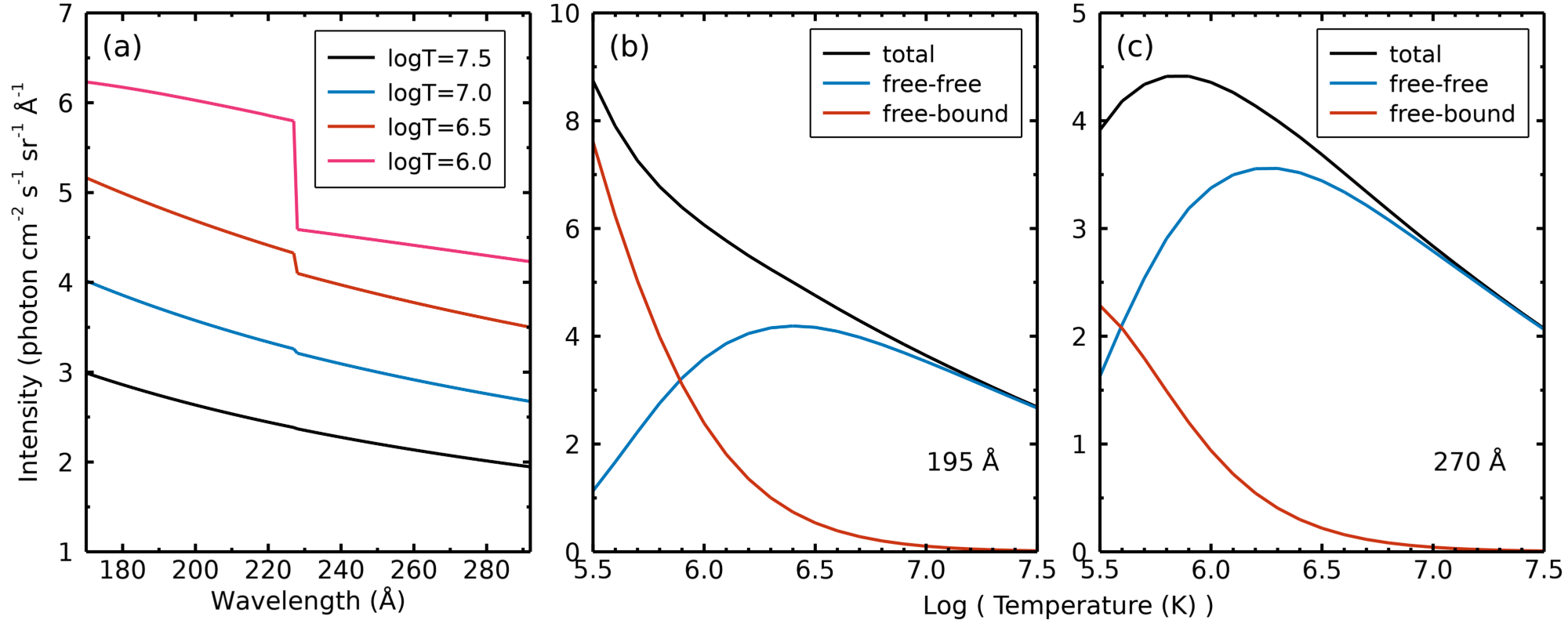}
    \caption{Panel (a) shows the continuum intensity calculated by the CHIANTI software for four temperatures as a function of wavelength. Panels (b) and (c) show the free-free continuum (blue), free-bound continuum (red) and total continuum (black) as a function of temperature, calculated at 195~\AA\ (b) and 270~\AA\ (c). }
    \label{fig:cont}
\end{figure}

The properties of the EUV continuum at the EIS wavelengths are illustrated in Figure~\ref{fig:cont} using data from CHIANTI. The continuum is produced by three processes: free-free (Bremsstrahlung), free-bound and two-photon, although the latter is negligible. The CHIANTI models were run using the photospheric abundances from \citet{2021A&A...653A.141A}, the default CHIANTI ionization balance, and a density of $10^{10}$~cm$^{-3}$.

Panel \ref{fig:cont}(a) shows the variation of the total continuum intensity with wavelength over the EIS wavelength range, for four isothermal temperatures. The continuum intensity is seen to increase with decreasing  temperature, hence it is critical to determine the plasma's temperature distribution using a DEM method to accurately model the continuum. The step jump in the continuum intensity at 227~\AA\ is due to the \ion{He}{ii} continuum edge. The step becomes larger at lower temperatures due to the increasing contribution of free-bound emission (Panels (b) and (c)).

Panels (b) and (c) show the relative contributions of free-free and free-bound emission to the total continuum at the wavelengths 195~\AA\ (Panel (b)) and 270~\AA\ (Panel (c)), which correspond to the peaks in the SW and LW effective area curves, respectively. These show that free-free dominates for typical flare temperatures of $10^7$~K, with free-bound becoming dominant at temperatures below $10^6$~K.

%================
\section{Dataset}\label{sec:dataset}

Data from the EIS instrument on the Hinode spacecraft are analyzed in this article. EIS is an imaging slit spectrometer that observes the two wavelength channels 171--212~\AA\ and 246--292~\AA\ that are referred to as the short-wavelength (SW) and long-wavelength (LW) channels, respectively. The spatial resolution is 3\as\--4\as\ and the spectral resolution is 3000--4000.

\begin{figure}
    \centering
    \includegraphics[width=\linewidth]{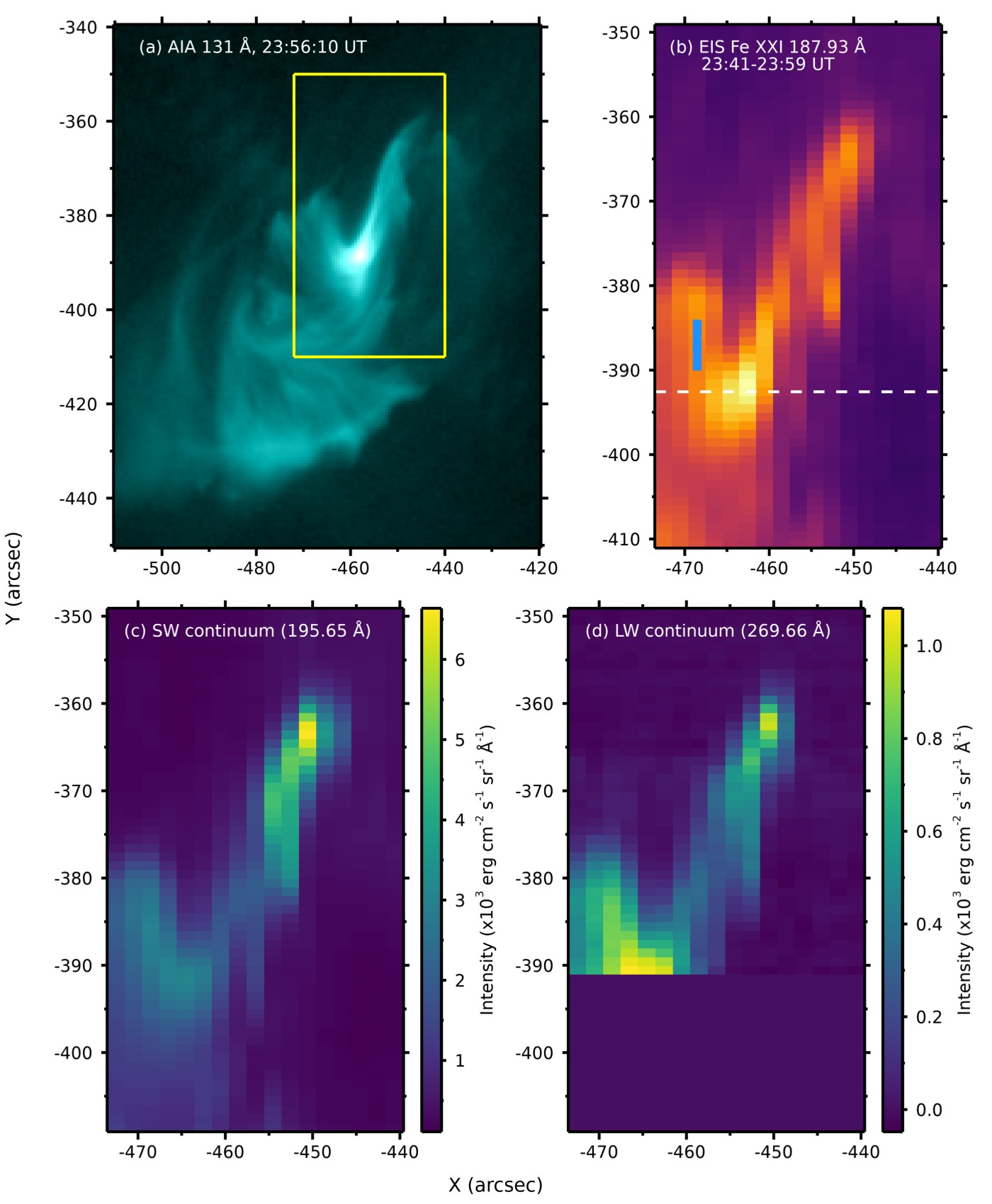}
    \caption{Panel (a) shows an AIA 131~\AA\ image from 23:56:10~UT. The yellow box highlights the sub-region shown in Panels (b)--(d). Panel (b) shows a section of the EIS raster image formed in the \ion{Fe}{xxi} 187.93~\AA\ line. Panels (c) and (d) show the same spatial region in the continuum at 195.65~\AA\ and 269.66~\AA, respectively. EIS rasters from right to left, and the exposures were taken between 23:41~UT and 23:59~UT. The dashed line on Panel (b) shows the approximate $y$ location at which the EIS LW channel is truncated such that no data for $y$-pixels below this line are available. The small blue squares on Panel (b) show the spatial locations used to create the continuum spectrum. Panels (a) and (b) are displayed with a cube-root intensity scaling, and Panels (c) and (d) are displayed with a linear intensity scaling.}
    \label{fig:context}
\end{figure}

Hinode Operation Plan (HOP) 460 was run in support of a Solar Orbiter Operation Plan to make observations when Solar Orbiter and Earth were at quadrature during 2024 September 30 to October 2. As part of HOP 460, 
EIS performed the study \textsf{Atlas\_60} that yielded a single raster scan between 22:57~UT and 23:59~UT that was centered on a filament to the north-west of active region AR 13842. However, it was a small filament within AR 13842 that erupted, and the corresponding flare was located in the bottom-left corner of the EIS raster. The flare had a GOES class of M7.7 and reached peak X-ray brightness at 23:59~UT. Figure~\ref{fig:context}(a) is an image from the AIA 131~\AA\ channel (dominated by \ion{Fe}{xxi} 128.75~\AA\ emission) at 23:56~UT, where a bright, U-shaped loop can be seen. Panel (b) is a section of the EIS raster image in the \ion{Fe}{xxi} 187.93~\AA\ line that corresponds to the yellow box on Panel (a). Panels (c) and (d) show images of the same region formed in the SW and LW continua.

The EIS raster uses the 2\as\ slit with a 2\as\ step size. The entire spectral range is downloaded, and the exposure time is 60~s. This is an unusually long exposure time for an EIS flare observation, and many of the most-commonly studied EIS emission lines are saturated in the flare. However, the spectrum reveals many weak flare emission lines and it also demonstrates strong continuum emission, which is the focus of the present work.

The location of the flare at the bottom of the EIS raster has the disadvantage that part of the brightest emission is not visible in the EIS LW channel. The two channels have their own detectors, and there is a physical offset between the two in the solar-$y$ direction of about 18 detector pixels (one pixel corresponds to 1\as\ angular resolution). Thus a solar feature found at $y$-pixel 1 on the LW detector would be found at $y$-pixel 19 in the SW channel. The dashed line on Figure~\ref{fig:context}(b) indicates the $y$-location below which there are no LW channel data, and Figure~\ref{fig:context}(d) shows the truncated image from the LW continuum. 

A single spectrum is used for measuring the continuum in this work, and exposure 58 was selected, indicated by the location of the small blue squares on Figure~\ref{fig:context}(b). This location is above the dashed line on Figure~\ref{fig:context}(b), hence the complete LW spectrum is available. The  loop top is brighter, but the LW \ion{Fe}{xxiii} and \ion{Fe}{xxiv} lines that are needed for the DEM analysis are saturated there. The loop leg and footpoint between $x=-460$ to $-450$ show non-Gaussian profiles related to chromospheric evaporation and the filament eruption, making measurements of the emission lines more difficult.

The continuum images in Figure~\ref{fig:context}(c) and (d) reveal that 
the loop footpoint at position ($-450$,$-362$) is significantly brighter in the SW continuum image compared to the looptop and leg in the region $x=-470$ to $-460$ and $y=-395$ to $-380$. For the LW image, the two regions are of comparable brightness, whereas for the SW image the footpoint is around a factor two brighter. The loop footpoint shows strong transition region emission with, for example, the \ion{O}{iv} lines at 272~\AA\ and 279~\AA\ comparable to the flare kernel spectrum presented by \citet{2024ApJ...966..102Y}. The strong transition region emission will lead to relatively more free-bound emission compared to free-free emission in the footpoint compared to the loop (see Figure~\ref{fig:cont}), and thus the SW continuum shows an enhancement compared to the LW channel due to the helium edge at 227~\AA. This is not discussed further here, but demonstrates that the EIS continuum contains valuable temperature information for flares. 

The September 30 flare has been studied in five other works, mostly focused on jets and nanojets observed during the filament eruption by the Extreme Ultraviolet Imager on Solar Orbiter \citep{2025ApJ...985L..12G, 2025ApJ...988L..65B,2025A&A...702A.189T,2026A&A...705A.113C,2025MNRAS.544.1758W}. None of these articles presented the EIS data from the flare.

%=========================
\section{Data preparation}\label{sec:prep}

\begin{figure}[t]
    \centering
    \includegraphics[width=0.7\linewidth]{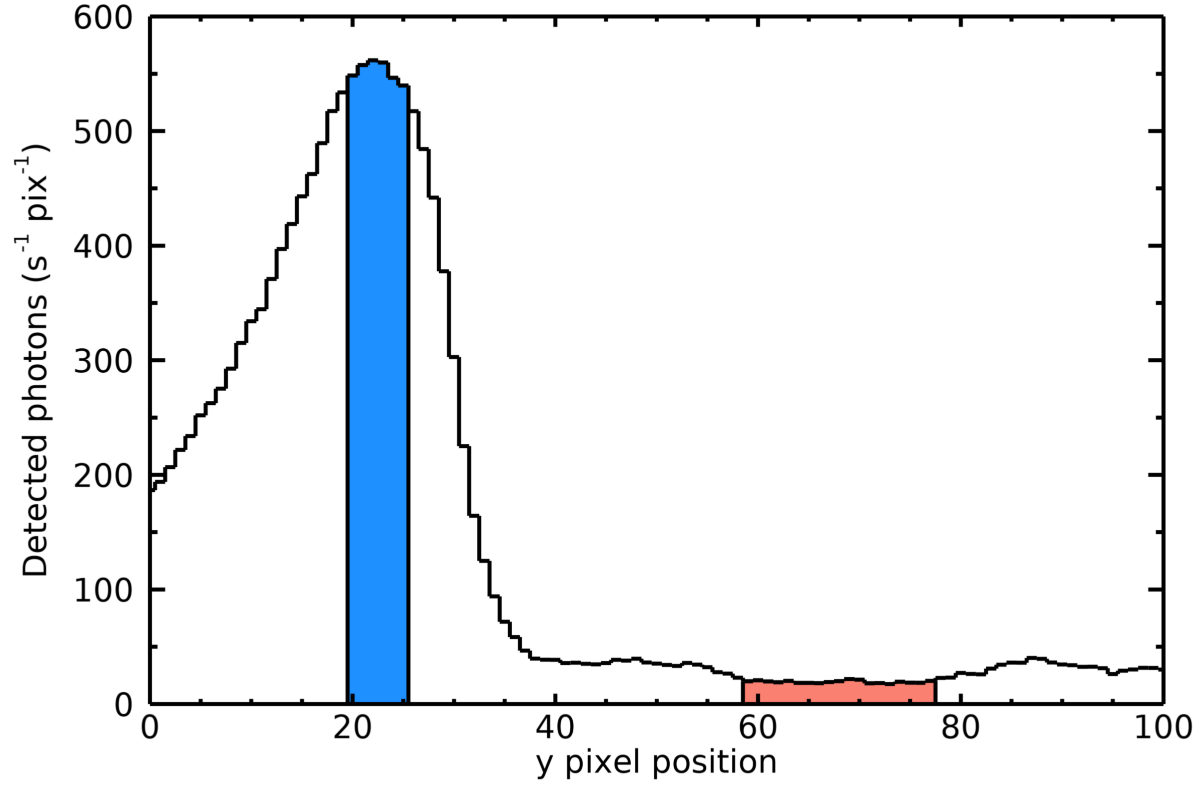}
    \caption{The variation of the continuum intensity at 195.79~\AA\ along the EIS slit for exposure number 58. The blue region highlights the pixels averaged to produce the flare spectrum, and the red region highlights the pixels averaged to produce the background spectrum.}
    \label{fig:cont-bg}
\end{figure}

The flare continuum emission is used to create new effective area curves for the SW and LW channels. Hence the dataset 
was calibrated using the \textsf{/photon} option in the call to \textsf{eis\_prep}, leading to intensity arrays in photon-event units, i.e., the radiometric calibration is not applied.
As discussed in the previous section, a small spatial region in exposure 58 of the raster was selected as the best location for the continuum analysis.
This exposure was obtained between 23:55:54~UT and 23:56:54~UT, and the flare spectrum was created with the routine \textsf{eis\_mask\_spectrum} by averaging six adjacent pixels in the $y$-direction that  are identified with small blue squares on Figure~\ref{fig:context}(b). This routine automatically corrects the EIS spectral tilt, whereby slit images in different lines are offset in the $y$-direction by as much as 20 pixels \citep{2009A&A...495..587Y}.

The continuum  to be measured from the EIS flare spectrum corresponds to those parts of the spectrum that are deemed to be free of emission lines. It comprises four components: (i) true continuum emission from the flare, (ii) true continuum from the background corona at the flare location, (iii) a pseudo continuum comprised of weak, unresolved emission lines, and (iv) the detector background emission. The pseudo continuum is addressed in Appendix~\ref{app:pseudo} where we estimate that it is around 5\%\ of the true continuum from the flare, and we do not consider it further in the analysis. 
The detector background can be measured from CCD dark frames and is found to be approximately uniform across the EIS detectors (Appendix~\ref{app:bg}). However, the background level varies with time and so it is necessary to directly estimate it from the dataset. For the present work, we select a low intensity region north of the flare to estimate both the true continuum from the background corona (ii) and the detector background (iv). We highlight that there is always a non-zero true continuum in the EIS spectra. In the quiet Sun it is at the level of around 1~photon in a 60~s exposure at 195~\AA, while in active regions it is around 20--30~photons.

To illustrate how the background is removed in the present case, consider Figure~\ref{fig:cont-bg}. A wavelength region of 7 pixels (0.156~\AA) centered on 195.79~\AA\ and free of emission lines is selected and the intensity averaged over wavelength. Figure~\ref{fig:cont-bg} plots this intensity as a function of $y$ for the flare exposure. The strong emission from the flare is clearly seen between pixels 0 and 35, and the six pixels averaged to yield the flare spectrum are denoted by the blue region. The intensity is much lower, but non-zero, for $y$-pixels 40 and higher. A minimum occurs around pixels 60--80, and this is assumed to represent the background. We therefore create another spectrum averaged over the 19 $y$-pixels identified by the red region in Figure~\ref{fig:cont-bg}. This is the background spectrum and is subtracted from the flare spectrum to yield a spectrum for which the continuum represents the true flare continuum. We refer to this spectrum as the ``photon spectrum" in the following sections.

\begin{figure}[t]
    \centering
    \includegraphics[width=\linewidth]{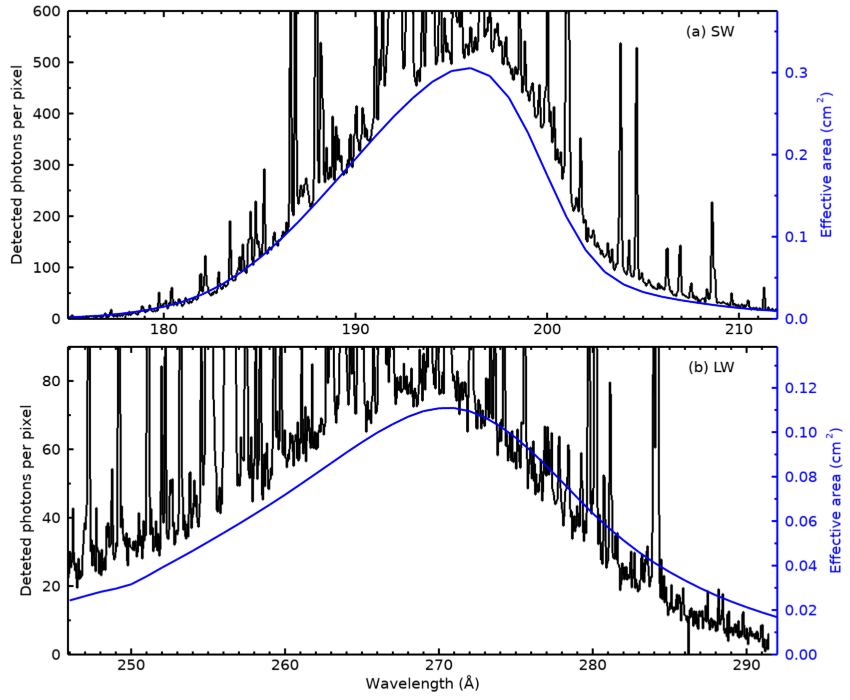}
    \caption{The black lines show the flare photon spectra for the two EIS channels. The spectra have been binned by two pixels, with the intensity averaged between the two bins. Some missing pixels in the SW spectrum around 192--193~\AA\ have been masked out for display purposes. The blue lines show the pre-launch effective area curves for the two channels. The scaling for these curves was chosen for display purposes and there was no attempt to fit the curves to the spectra.}
    \label{fig:ea-spec}
\end{figure}

Figure~\ref{fig:ea-spec} shows the photon spectra for the SW and LW channels, scaled so as to emphasize the continuum part of the spectra. The pre-launch effective area curves are over-plotted, showing the general shapes are in good agreement with the continuum, but there is an impression of wavelength-dependent changes to both curves. This comparison does not take account of the intrinsic wavelength dependence of the continuum, which is modeled correctly in the present work. The ratio of the continuum peak in the SW channel to that in the LW channel is around 6, whereas the ratio of the peaks of the effective area curves is 2.7. This is consistent with the findings of \citet{2013A&A...555A..47D}, \citet{2014ApJS..213...11W} and DZWW25 that the LW channel has degraded more than the SW channel by around a factor two. 

As part of the analysis procedure described in the following section it is necessary to use the flare emission line intensities to obtain a DEM curve. For this, the pre-launch radiometric calibration was applied, leading to units of \ecssa. The same six spatial pixels used to define the continuum spectrum were averaged, but the background spectrum was not subtracted. This spectrum is referred to as the ``calibrated spectrum" below.

%==================
\section{Analysis Procedure}\label{sec:proc}

The goal of this project is to use the flare continuum measurement to obtain effective area curves for the EIS SW and LW channels. The procedure is as follows:
\begin{enumerate}
    \item Measure continuum intensities from the photon spectrum at sets of wavelengths across both wavelength channels.
    \item Measure a set of emission line intensities from the LW calibrated spectrum.
    \item Perform a DEM analysis using the LW emission line intensities.
    \item From the DEM, create a model for the continuum emission in both the SW and LW channels.
    \item Derive updated SW and LW effective area curves by comparing the continuum intensities with the continuum model.
    \item With the modified effective area curves, adjust the LW emission line intensities, and repeat steps (3) to (5).
\end{enumerate}
The DEM analysis is applied only to the LW channel lines for two reasons. Firstly, the method takes the pre-launch effective areas as a starting point, and the previous works have demonstrated that the two channels have degraded at different rates. Thus a DEM calculated from a mixture of SW and LW lines would not be accurate. Secondly, the hottest ion available for the DEM analysis is \ion{Fe}{xxiv}, which is needed to constrain the high temperature part of the DEM. This ion has lines in both channels at 192.03~\AA\ and 255.11~\AA, but the former is saturated in the flare spectrum and cannot be used.

%---------------------------------
\subsection{Continuum measurement}\label{sec:cont-meas}

\begin{figure}[t]
    \centering
    \includegraphics[width=\linewidth]{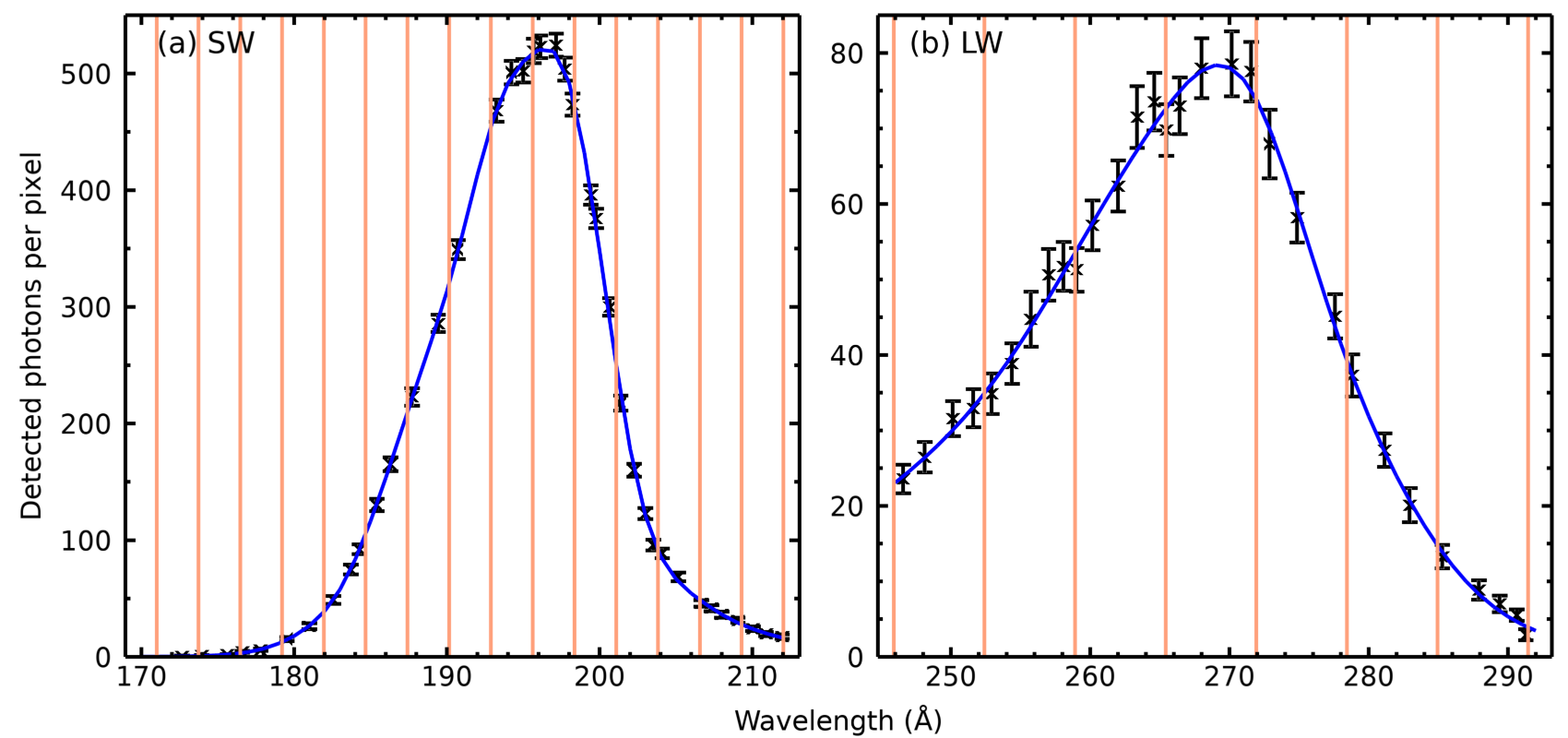}
    \caption{The crosses with error bars show the measured continuum intensity values for the EIS SW (a) and LW (b) channels. Spline fits to the intensities are plotted with blue lines, which are defined on a wavelength spacing of 1~\AA. The locations of the spline nodes are indicated with vertical red lines.}
    \label{fig:cont-fit}
\end{figure}

The continuum was measured at multiple wavelength locations in both the SW and LW channels. The photon spectra were input to the routine \textsf{spec\_gauss\_eis}, which allows the user to graphically zoom in to a section of the spectrum. Small sections were selected that were deemed to be free of emission lines, and the average intensities and uncertainties in each section were computed. The sizes of the sections ranged from 5  to 22 pixels (0.12 to 0.51~\AA). The resulting continuum intensities for the two channels are shown in Figure~\ref{fig:cont-fit}.

The continuum intensities for a channel were fit with a function of the form $\exp(f(\lambda;a))$, where $f$ is a cubic spline and $a$ the function values at a set of $n$ node points that span the EIS channel's wavelength range. The nodes are equally spaced across the range, and $n=16$ and 8 for the SW and LW channels, respectively. A larger number of nodes were used for the SW channel due to the much larger variation in effective area. The node locations are indicated with the red vertical lines on Figure~\ref{fig:cont-fit}.
The spline is defined with the IDL functions \textsf{spl\_init} and \textsf{spl\_interp}, and fitting is performed with the \textsf{mpfit} routines \citep{2009ASPC..411..251M}.

%------------------------
\subsection{Emission Line Measurements}\label{sec:line-meas}

Figure~\ref{fig:cont}(a) shows that the continuum intensity is relatively stronger at lower temperatures compared to typical flare temperatures of $10^7$~K. To model the continuum accurately it is therefore important to model the temperature distribution of the plasma. Here a DEM  curve is derived using only emission lines from the EIS LW channel. 

\begin{deluxetable}{ccccccccc}[t]
\tabletypesize{\scriptsize}
\tablecaption{Emission lines used for the DEM analysis. \label{tbl:dem}}
\tablehead{
  &\multicolumn{2}{c}{Wavelength (\AA)}
  &&\multicolumn{2}{c}{Intensity (\ecss)} \\
  \cline{2-3} \cline{5-6}
  \colhead{Ion} &
  \colhead{CHIANTI} &
  \colhead{Measured} &&
  \colhead{Observed} &
  \colhead{Model} &
  \colhead{Ratio\tablenotemark{a}} &
  \colhead{$\log\,T_\mathrm{max}$} &
    \colhead{$\log\,T_\mathrm{eff}$}
} 
\startdata
  \ion{Si}{vii} &    275.368 &    275.405 && $  108.9 \pm    21.3$ &   106.1 & $   0.97 \pm    0.19$ &   5.75 &   5.75\\
    \ion{Si}{x} &    258.374 &    258.399 && $ 1125.2 \pm   170.3$ &  1037.0 & $   0.92 \pm    0.14$ &   6.15 &   6.25\\
  \ion{Fe}{xiv} &    264.788 &    264.814 && $ 4438.1 \pm   666.6$ &  4643.1 & $   1.05 \pm    0.16$ &   6.25 &   6.25\\
  \ion{Fe}{xiv} &    274.203 &    274.226 && $ 2305.0 \pm   346.6$ &  1722.3 & $   0.75 \pm    0.11$ &   6.25 &   6.25\\
   \ion{Fe}{xv} &    284.163 &    284.190 && $35856.0 \pm  5381.5$ & 42325.5 & $   1.18 \pm    0.18$ &   6.35 &   6.55\\
  \ion{Fe}{xvi} &    262.976 &    263.010 && $13242.0 \pm  1987.1$ &  9356.2 & $   0.71 \pm    0.11$ &   6.45 &   6.85\\
 \ion{Fe}{xvii} &    269.420 &    269.443 && $ 1162.9 \pm   175.5$ &  1215.5 & $   1.05 \pm    0.16$ &   6.75 &   6.85\\
   \ion{Ti}{xx} &    259.272 &    259.307 && $  610.1 \pm    93.7$ &   510.8 & $   0.84 \pm    0.13$ &   6.95 &   7.05\\
  \ion{Fe}{xxi} &    270.546 &    270.614 && $ 1273.1 \pm   207.3$ &  1262.9 & $   0.99 \pm    0.16$ &   7.05 &   7.05\\
 \ion{Fe}{xxii} &    247.188 &    247.234 && $ 2599.7 \pm   393.7$ &  2792.4 & $   1.07 \pm    0.16$ &   7.15 &   7.15\\
\ion{Fe}{xxiii} &    263.765 &    263.783 && $ 9535.9 \pm  1431.7$ &  7900.0 & $   0.83 \pm    0.12$ &   7.15 &   7.15\\
 \ion{Fe}{xxiv} &    255.113 &    255.134 && $19974.0 \pm  2997.7$ & 23492.8 & $   1.18 \pm    0.18$ &   7.25 &   7.15\\

\enddata
\tablenotetext{a}{Ratio of model to observed intensity.}
\end{deluxetable}

Good coverage of the hottest plasma is available through lines of \ion{Fe}{xxi--xxiv} and \ion{Ti}{xx}. At lower temperatures the coverage is less good, but sufficient to constrain the DEM curve down to $\log\,T=5.8$, where \ion{Si}{vii} is formed. The LW channel contains lines from cooler ions such as \ion{O}{iv} 279.93~\AA\ and \ion{Mg}{v} 276.58~\AA, but these lines cannot be measured implying the emission measure is very low at these temperatures. 

The emission lines in the calibrated spectrum were fit with Gaussian functions using the routine \textsf{spec\_gauss\_eis}. All of the lines except \ion{Fe}{xxi} 270.55~\AA\ could be measured easily by fitting a single Gaussian function. The \ion{Fe}{xxi} line is blended with \ion{Fe}{xiv} 270.52~\AA, leading to an asymmetric intensity feature in the spectrum. It was fit simultaneously with \ion{Fe}{xiv} 274.20~\AA\ by requiring the two \ion{Fe}{xiv} lines to have the same width, but otherwise the parameters of the three Gaussians were free to vary. 

Typical electron number densities ($N_\mathrm{e}$) in the flaring corona are $\log\,(N_\mathrm{e}/\mathrm{cm}^{-3})=10$ to 12 and
all lines in Table~\ref{tbl:dem} show a weak density dependence over this range except for the two \ion{Fe}{xiv} lines.
However the sum of the lines' contribution functions is insensitive to density and so they were included in the DEM analysis as a single line by summing their intensities and contribution functions.
Table~\ref{tbl:dem} lists the lines separately, and the model-to-observed ratio for the combined lines is $0.84\pm 0.09$. The use of density-insensitive lines for the DEM analysis ensures that the solution is not dependent on the choice of density. The density used here is discussed in the following section.

Table~\ref{tbl:dem} gives the observed intensities obtained with the final effective area curves, and the model intensities are those obtained from the final DEM solution. The temperature of maximum ionization, $T_\mathrm{max}$, is the temperature at which the ion's ionization balance curve peaks, and the effective temperature, $T_\mathrm{eff}$, is obtained by multiplying the line's contribution function with the DEM curve and finding the maximum. It is thus the temperature that gives the largest component to the line's emission. 

%------------------------
\subsection{DEM analysis}\label{sect:dem}

Solving for a DEM requires specifying an element abundance file, an ionization equilibrium file, and a  constant density or pressure through the atmosphere. All of the lines used for the DEM analysis belong to elements with a low\footnote{The boundary between low and high FIP is typically set at 10~eV \citep[e.g.,][]{2012ApJ...755...33S}, with elements such as Mg, Si and Fe considered low-FIP and elements such as H, O and Ne considered high-FIP.} first ionization potential (FIP), and abundance anomalies related to the FIP effect are not relevant. However, the dominant contribution to the continuum comes from hydrogen, which is considered a high FIP element. It is thus likely that the continuum to line ratio will be anomalous if the FIP effect is operating. The DEM was initially generated with the photospheric abundances of \citet{2021A&A...653A.141A}. The continuum intensity predicted from the DEM analysis was found to be too weak compared to the emission lines, and so the FIP bias was adjusted. That is, the abundances of the low FIP elements were reduced, keeping the high FIP element abundances the same. The FIP bias was adjusted manually until a good fit to both the emission line intensities and the continuum was found. This method of obtaining the FIP bias is similar to that applied to spectra from the EUV Variability Experiment on SDO  by \citet{2014ApJ...786L...2W}. The final FIP bias was 0.57, which is an example of the unusual inverse-FIP effect \citep[e.g.,][]{2015ApJ...808L...7D}, which is discussed further in Section~\ref{sec:xsm}.

The plasma density was estimated using the \ion{Ca}{xv} 181.90~\AA/182.86~\AA\  ratio as this is one of the hottest diagnostics available to EIS, and the lines are close in wavelength so the ratio is not sensitive the radiometric calibration. The observed ratio using the pre-launch calibration is $2.06\pm 0.25$, giving a density of $\log\,(N_\mathrm{e}/\mathrm{cm}^{-3})=10.66^{+0.08}_{-0.09}$. \ion{Ca}{xv} has  $\log\,(T_\mathrm{max}/\mathrm{K})=6.65$ and so a pressure, $P$, of $2.04\times 10^{17}$~K\,cm$^{-3}$ is assumed for the DEM analysis. Two further density diagnostics are available: \ion{Si}{x} 258.37/261.06 and \ion{Fe}{xiv} 264.79/274.20, formed at $\log\,(T/\mathrm{K})=6.15$ and 6.30, respectively. Using the final effective area curves (Section~\ref{sec:cont-calc}) these ratios yield densities of $\log\,(N_\mathrm{e}/\mathrm{cm}^{-3})=10.0$ and 10.1, respectively, somewhat lower than for \ion{Ca}{xv}.
As the continuum intensity is independent of density, a precise pressure is not critical to the analysis and we use the pressure derived from \ion{Ca}{xv} since the formation temperature is closer to that of the hot flare lines.
The ionization balance is calculated with the CHIANTI software using the measured pressure value.

The temperature range used for the DEM calculation is calculated by the routine \textsf{ch\_dem\_get\_ltemp} by considering the temperature range for each ion for which the contribution function is greater than 0.005 of the function's peak.  This yielded a temperature range of $\log\,(T/\mathrm{K})=5.45$ to 8.05 in 0.1~dex intervals. The upper limit is very high due to the high-temperature tail of the \ion{Fe}{xxiv} contribution function. Since the tail is not constrained by any other species, the upper limit of the DEM temperature range was restricted to $\log\,(T/\mathrm{K})=7.55$. 

The uncertainties on the measured line intensities come from photon noise and fitting uncertainties. Due to the high signal in the flare, the uncertainties are small, ranging from 0.4\%\ to 13\%\ of the line intensity. It is a standard procedure in DEM analyses to insert an artificial uncertainty of around 15--20\%\ of the measured intensity that represents uncertainties due to atomic data and the instrument calibration---see Section~3.1.1 of \citet{2019ApJ...885....7K}, for example. For the present work a 15\%\ uncertainty is added in quadrature to the measurement uncertainty, giving the uncertainties in Table~\ref{tbl:dem}.

\begin{figure}[t]
    \centering
    \includegraphics[width=\linewidth]{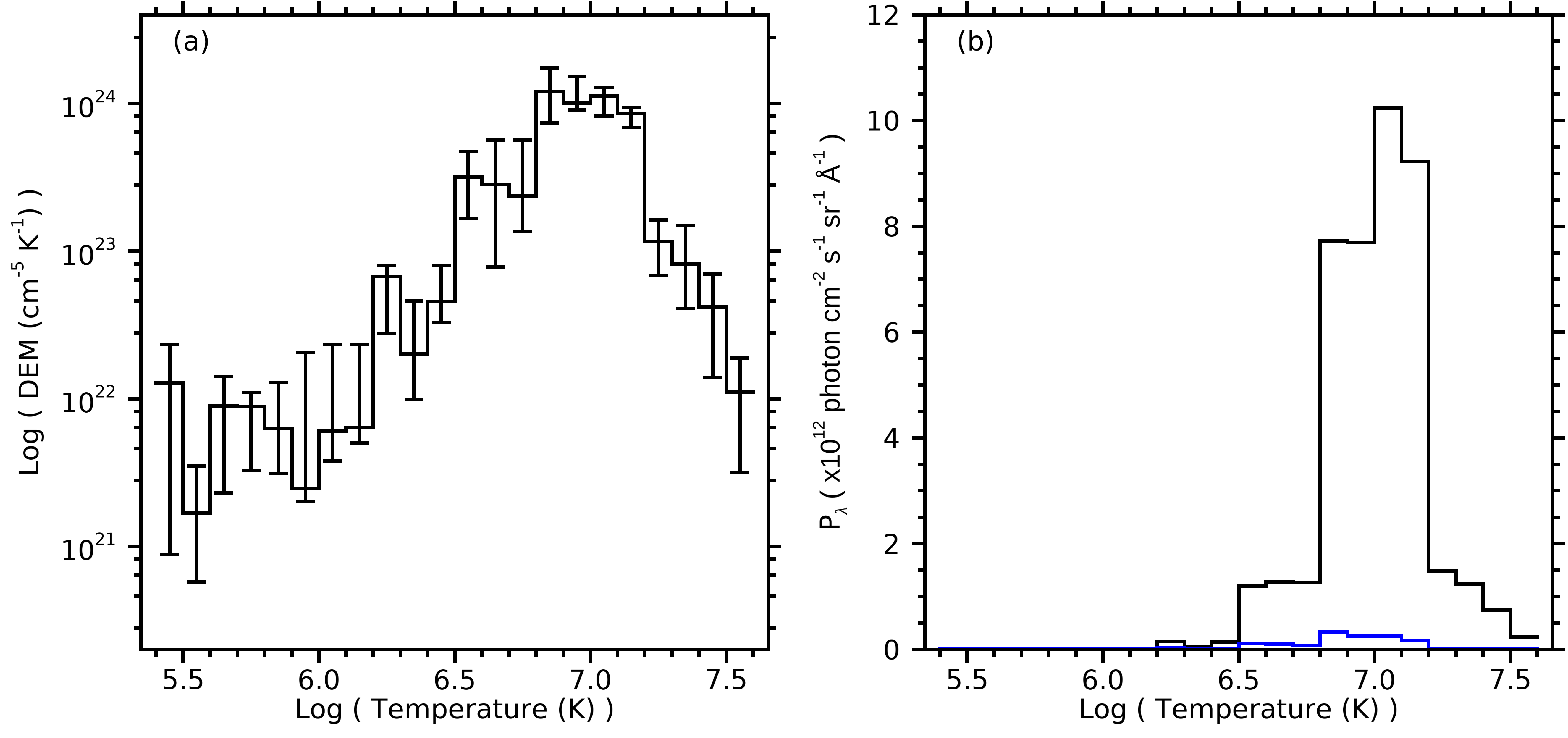}
    \caption{(a) The DEM obtained with the MCMC method using emission line intensities computed with the new effective area curves. (b) Plot showing the temperatures at which the continuum is formed at a wavelength of 195.79~\AA. The blue line shows the free-bound component.}
    \label{fig:dem}
\end{figure}

Figure~\ref{fig:dem}(a) shows the final DEM distribution from which the new effective area curves were derived. Convolving with the continuum emissivity function gives the curve in Figure~\ref{fig:dem}(b), showing that the continuum is dominated by plasma around $\log\,T=7.0$: 82\%\ of the continuum emission comes from the temperature range $\log\,(T/\mathrm{K})=6.85$ to 7.15. The blue curve on Figure~\ref{fig:dem}(b) shows the free-bound continuum contribution is negligible, as expected given the different temperature dependencies of the free-free and free-bound continua (Figures~\ref{fig:cont}(b) and (c)).
The nature of the MCMC method means that the code generates a slightly different DEM solution each time, but the general shape remains the same and the variations do not lead to any significant changes to the final effective area solution since the continuum is insensitive to small changes in the temperature distribution. 

A fundamental assumption in the DEM analysis is that the plasma is in ionization equilibrium, and hence the CHIANTI emission line modeling software  can be applied. Flares are very dynamic and non-equilibrium conditions can be expected in the impulsive phase when the chromospheric plasma is heated rapidly and strong plasma flows develop. The EIS exposure was obtained two minutes before the time of peak X-ray flux of the flare, at which point the flare loops have filled with hot, dense plasma and can be expected to be in an equilibrium state. The high density found from the \ion{Ca}{xv} diagnostic supports this statement, since electron collisions quickly drive a plasma to equilibrium at high density. The emission line profiles are also well fit with Gaussian functions, further supporting the absence of non-equilibrium effects.

The DEM analysis also assumes the emission lines are optically thin. If they are optically thick, then the shape and magnitude of the derived DEM could be affected. Since the DEM is dominated by plasma near 10~MK (Figure~\ref{fig:dem}), opacity in the hot iron lines is the central concern. The lines of \ion{Fe}{xxi--xxiii} are all intercombination transitions with small oscillator strengths, hence we consider \ion{Fe}{xxiv} 255.11~\AA, a resonance line. Using Equation~6 of \citet{2011ApJ...740...70M}, the \ion{Ca}{xv} density derived earlier, and the $T_\mathrm{max}$ value from Table~\ref{tbl:dem} we obtain an opacity, $\tau_0$, of $2.7\times 10^{-3}$ ($h$/Mm), where $h$ is the plasma column depth. A line is considered optically thick if $\tau_0 \ge 1$, which requires $h\ge 650$~Mm, much larger than the size of the flare loop ($\approx$30~Mm), hence  optical depth effects can be ignored for this line. The same calculation for the cooler resonance lines \ion{Fe}{xiv} 274.20~\AA\ and \ion{Fe}{xv} 284.16~\AA\ yields column depths at $\tau_0 =1$ of 6~Mm and 20~Mm, and thus there may be weak opacity effects in these lines. This would lead to a small underestimate in the DEM around $\log\,(T/\mathrm{K})=$6.1--6.4, but since the DEM is more than an order magnitude weaker here than at $\log\,(T/\mathrm{K})=7.0$, this will have a negligible impact on the continuum analysis.

%---------------------------------
\subsection{Continuum calculation}\label{sec:cont-calc}

After the DEM has been calculated, a synthetic continuum spectrum is created with the CHIANTI software using the DEM and the DEM input parameters (ionization balance file, abundance file and the constant pressure value). This yields a spectrum, $P_\lambda$, in units of \pcssa. This is converted to a predicted EIS count rate, $C_\lambda$, with the expression
\begin{equation}
    C_\lambda = p\,S\,t_\mathrm{exp}\,P_\lambda \,A_\mathrm{eff}(\lambda)
\end{equation}
where $p$ is the EIS pixel size in steradian ($2.349\times 10^{-11}$~sr for a 1\as\ square pixel), $S$ is the EIS slit width in arcsec, $t_\mathrm{exp}$ the exposure time in seconds, and $A_\mathrm{eff}$ the EIS effective area. The latter is the effective area used to generate the emission line intensities used for the DEM analysis.

A new estimate of $A_\mathrm{eff}$ is made by dividing the observed continuum intensity (Section~\ref{sec:cont-meas}) by $C_\lambda$. The SW and LW curves are normalized to the DZWW25 curves by requiring that $A_\mathrm{eff}(195.12\,\mathrm{\AA})$ takes the same value, where the DZWW25 value is that which applied on 2022 April 1, the most recent date for which the curves were calculated. 

The new effective area curve is used to modify the emission line intensities used as input to the DEM code, which then leads to a new DEM curve and hence a new continuum spectrum. This process could be repeated multiple times to refine the effective area curve, but it was found that a second iteration led to negligible changes ($<1$\%) to $A_\mathrm{eff}$ and so the result from the first iteration  was used as the final effective area solution. Although the change from the pre-launch effective area to the new effective area is very significant, it did not change the basic shape of the DEM (Figure~\ref{fig:dem}). In particular, the peak remained close to $\log\,(T/\mathrm{K})=7.0$, which dominates the contribution to the continuum.

%================
\section{Effective Area Results}\label{sec:results}

In this section comparisons of the new effective area (EA) curves with those of DZWW25 are performed. DZWW25 provided time-dependent curves over the EIS mission lifetime up to 2022 April. For subsequent dates, the curves are assumed not to change. The method for determining the EIS effective areas developed by DZWW25 consisted of three steps. (1) a DEM study of seven off-limb quiet Sun datasets that yielded relative EA curves at seven dates during the mission; (2) the application of insensitive line ratios to many quiet Sun and active region datasets to refine the relative EA curves at intervals of 6 months during the mission; and (3) the derivation of absolute EA curves by comparing count rates to those from the AIA 193~\AA\ channel.

\begin{figure}[t]
    \centering
    \includegraphics[width=\linewidth]{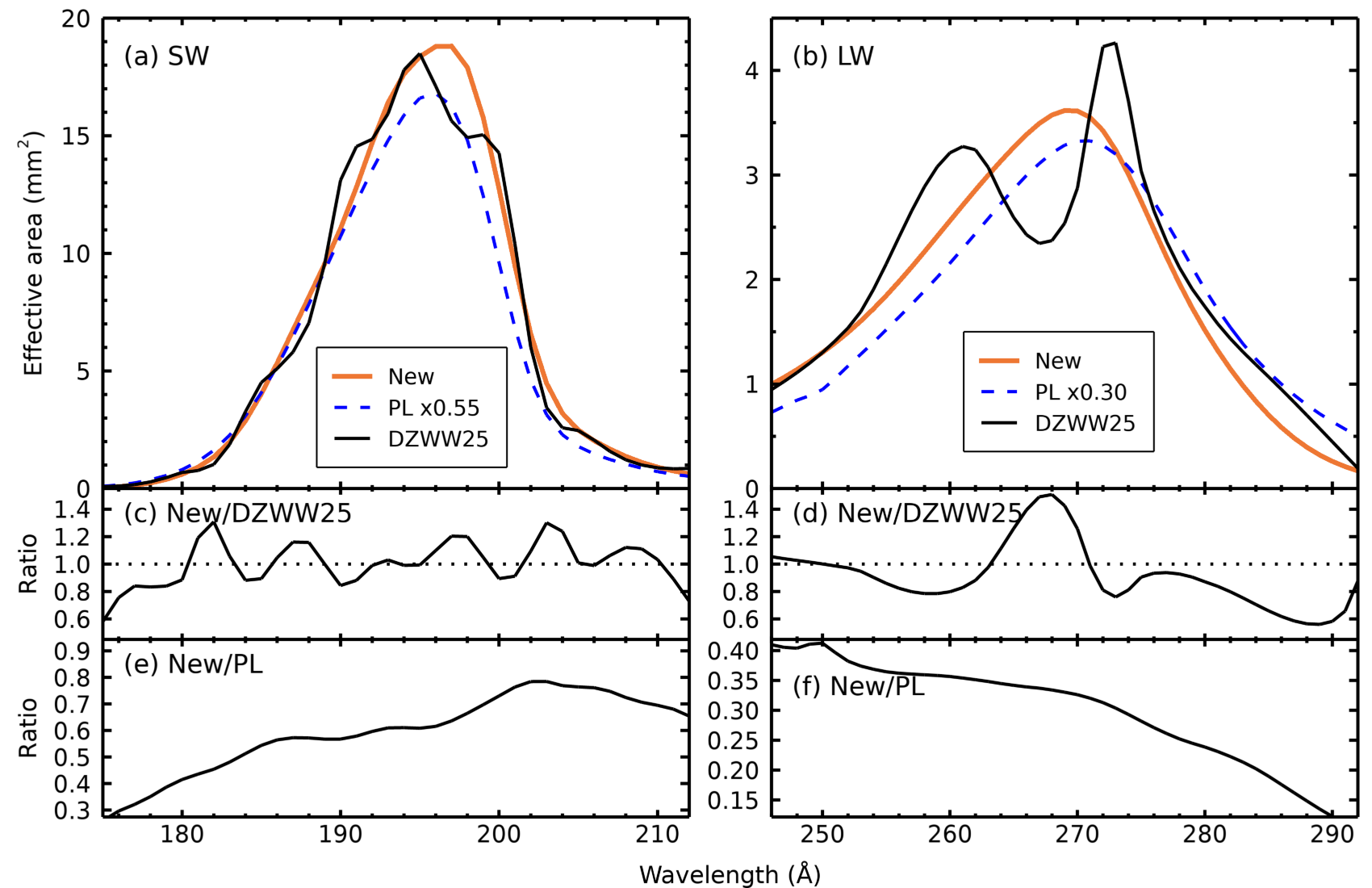}
    \caption{Panels (a) and (b) compare the new effective area curves (orange) with the DZWW25 curves (black) and the pre-launch curves (blue, dashed) for the SW and LW channels, respectively. The pre-launch curves have been multiplied by factors of 0.55 (SW) and 0.30 (LW) for display purposes. Panels (c) and (d) give the ratios of the new and DZWW25 curve pairs, with the new curve as the numerator. Panels (e) and (f) give the ratios of the new and pre-launch curve pairs. }
    \label{fig:ea-comp}
\end{figure}

Figure~\ref{fig:ea-comp} compares the new effective area curves with those from DZWW25. In Section~\ref{sec:insens} we check how well the continuum-derived EA curves reproduce the insensitive line ratios identified by DZWW25.
The lines from \ion{S}{x} are a special case due to the anomaly identified by DZWW25 for datasets obtained in 2021 and 2022, and are discussed in Section~\ref{sec:s10}.

The DZWW25 comparison for the  SW channel demonstrates agreement within $\pm 30$\%\ over the range 180--210~\AA\ (Figure~\ref{fig:ea-comp}(c)). There is much more structure in the DZWW25 curve, but this is not due to a greater number of node points used for the spline fit, since DZWW25 used 10 points compared to 16 here.

For the LW channel, the overall magnitudes of the EA curves agree quite well, showing that the two completely independent methods are consistent in measuring the relative degradation of the SW and LW channels. However, there is a large disagreement in the 255--275~\AA\ range, where the DZWW25 curve has a double peak but the present curve has a single peak. The largest difference is at 268~\AA\ where the present EA value is 51\%\ higher than the DZWW25. The differences in the curves in this region are reflected in some of the insensitive line ratios discussed in Section~\ref{sec:insens}.

Figure~\ref{fig:ea-comp} also compares the new effective area curves with the original, pre-launch curves. The ratio curves in Panels (e) and (f) are much smoother than for the DZWW25 comparisons (Panels (c) and (d)) suggesting that localized changes in the curves have been small. Within the LW channel, sensitivity has declined most at long wavelengths, which is consistent with the build up of a thin layer of carbon on the detectors \citep{2013SoPh..282..629M} and would also explain the smaller sensitivity decline of the SW channel. However, within the SW channel there is an opposite trend of decreasing sensitivity with decreasing wavelength below 200~\AA, suggesting an additional, unknown factor affecting the EIS sensitivity in this range.

\begin{deluxetable}{lll}[t]
\tabletypesize{\normalsize}
\tablecaption{Observations used for insensitive line ratio measurements. \label{tbl:insens}}
\tablehead{
  \colhead{Date ID} & 
  \colhead{Study} &
  \colhead{Comment}
} 
\startdata
20240913\_002013 & \textsf{Atlas\_60} & Fan loop (\ion{Fe}{viii--x}) \\
20240930\_225718 & \textsf{Atlas\_60} & Flare loop (\ion{Fe}{xvi--xvii}) \\
20241003\_115349 & \textsf{Atlas\_120} & Quiet Sun (\ion{S}{x}) \\
20241014\_001949 & \textsf{FlareResponse04} & Flare loop (\ion{Fe}{xxiv}) \\
20241020\_004431 & \textsf{Atlas\_60} & Off-limb loop (\ion{Fe}{xi--xiv}, \ion{Si}{x}) \\
\enddata
\end{deluxetable}

%---------------------------------
\subsection{Line ratio comparison}\label{sec:insens}

Table~1 of DZWW25 gives a list of emission line ratios that the authors used to help determine the EIS calibration. In this section we measure these ratios from
spectra obtained close in time to the 2024 September 30 observation using the new calibration, and compare with the theoretical ratios and the ratios obtained from the DZWW25 calibration.

Five datasets were used for measuring the ratios, and are listed in Table~\ref{tbl:insens}. The date ID gives the date and time of the start of each observation, for example, \textsf{20240913\_002013} corresponds to 00:20:13~UT on 2024 September 13. The datasets were chosen by assessing which types of observations would give the best measurements of line ratios. Thus, fan loops show strong emission from \ion{Fe}{viii--x}, and so the 20240913\_002013 observation was selected as it contains strong fan loop emission. Since the \ion{Fe}{xxiv} flare lines are saturated in the 2024 September 30 dataset, the flare observation from 2024 October 14 was used. The \ion{Fe}{xvi} and \ion{Fe}{xvii} lines were measured from the 2024 September 30 dataset, but from exposure 56 rather than exposure 58, which was used for the continuum  measurement. Exposure 56 is closer to the loop top and has stronger lines than exposure 58. A quiet Sun dataset was chosen specifically for \ion{S}{x} due to blending issues in active region spectra.

The theoretical ratios in Table~\ref{tbl:ratios} are obtained from CHIANTI 11.0 \citep{2024ApJ...974...71D} and are given in photon units. Some line pairs correspond to branching ratios that have a fixed ratio in all solar conditions. For the remaining line pairs the ratios are computed over the temperature range $\log\,T_\mathrm{max}\pm 0.15$, and the density range $\log\,N_\mathrm{e}=8.5--9.5$. The latter range is based on results for density diagnostics in the fan loop and off-limb loop spectra. The \ion{Mg}{vii} 280.75/276.15 ratio gives a density of $\log\,N_\mathrm{e}=9.2$ for the fan loop spectrum, and the \ion{Si}{x} 258.37/261.06, \ion{Fe}{xii} 186.88/195.12 and \ion{Fe}{xiv} 264.78/274.20 ratios give $\log\,N_\mathrm{e}=8.8-9.0$ for the off-limb loop spectrum. All of the ratios for the quiet Sun spectrum and the two flare spectra are insensitive to density. 
Table~\ref{tbl:ratios}  gives the minimum and maximum ratio values over the temperature and density ranges. The \ion{Fe}{xi} 178.06/182.17 ratio is omitted from the table as the 178.06~\AA\ line could not be measured in the off-limb loop spectrum, and the \ion{S}{xi} 285.59/281.40 branching ratio is omitted due to a known problem with the 285.59~\AA\ line \citep{1998A&A...329..291Y}. Three ratios of \ion{Fe}{viii} not considered by DZWW25 are listed in Table~\ref{tbl:ratios} as the lines are strong in the cool loop spectra and provide a valuable check over a wavelength region where the EA shows a large change. The \ion{Fe}{xiv} 270.52/274.20 ratio is preferred here over the 270.52/(264.78+274.20) ratio used by DZWW25 due to the problems noted above for the 255--275\,\AA\ region, and is discussed in the following section.

There is generally very good agreement between the ratios calculated with the two calibrations. Triangles on the right side of Table~\ref{tbl:ratios} indicate ratios for which the present and DZWW25 calibrations differ by more than 30\%. Unfilled triangles indicate that both ratios are still within the theoretical range of the ratio, or within $\pm 20$\,\%\ for branching ratios. The \ion{Fe}{x} 174.53/184.54, \ion{Fe}{xiii} 201.13/197.43 and \ion{Fe}{xiv} 289.15/274.20 ratios show significantly worse agreement with theory compared to DZWW25. The 174.53~\AA\ and 289.15~\AA\ lines are located where the continuum is very weak (Figure~\ref{fig:ea-spec}), and so the two ratios suggest the measured continuum intensities are under-estimates at these locations. We therefore caution that the continuum-derived effective area curves may not be accurate in the low EA regions $<$180~\AA\ and $>$285~\AA. The discrepancy for \ion{Fe}{xiii} 201.13/197.43 is likely due to a problem with the 197.43~\AA\ line as other ratios involving the 201.13~\AA\ are in good agreement with theory.

\begin{deluxetable}{lccccc}[t]
\tabletypesize{\small}
\tablecaption{Insensitive line ratios. \label{tbl:ratios}}
\tablehead{
  &&\multicolumn{3}{c}{Line ratio (photon units)}\\
  \cline{3-5}
  &Ratio wavelengths &&\multicolumn{2}{c}{Observed}\\
  \cline{4-5}
  \colhead{Ion} & 
  \colhead{(\AA)} &
  \colhead{Theory} & 
  \colhead{Present} & 
  \colhead{DZWW25}
} 
\startdata
\ion{Fe}{viii}\tablenotemark{a} & 194.66/185.21 & 0.24--0.26 &    0.25 &    0.27 \\ 
\ion{Fe}{viii}\tablenotemark{a} & 194.66/186.60 & 0.34--0.36 &    0.36 &    0.31 \\ 
\ion{Fe}{viii}\tablenotemark{a} & 185.21/255.35 & 5.05--9.76 &    6.33 &    6.89 \\ 
\ion{Fe}{ix} & 189.94/197.85 & 1.03--1.91 &    1.50 &    1.05 & $\triangleleft$ \\ 
\ion{Fe}{ix} & 188.49/197.85 & 1.87--3.20 &    2.46 &    2.22 \\ 
\ion{Fe}{x} & 174.53/184.54 & 3.92--4.56 &    6.15 &    4.01 & $\blacktriangleleft$\\ 
\ion{Fe}{x} & 177.24/184.54 & 2.29--2.61 &    2.85 &    2.77 \\ 
\ion{Fe}{x} & 190.04/184.54 & 0.34 &    0.35 &    0.34 \\ 
\ion{Fe}{x} & 207.45/184.54 & 0.13--0.25 &    0.15 &    0.19 \\ 
\ion{Fe}{x} & 257.26/184.54 & 0.69--1.65 &    1.12 &    1.03 \\ 
\ion{Fe}{xi} & 188.22/192.83 & 4.67 &    2.68 &    2.95 \\ 
\ion{Fe}{xi} & 202.71/188.30 & 0.145 &   0.103 &   0.116 \\ 
\ion{Fe}{xi} & 180.40/188.22 & 1.84--2.02 &    2.33 &    2.05 \\ 
\ion{Fe}{xi} & 257.55/188.22 & 0.144--0.303 &   0.220 &   0.153 & $\triangleleft$ \\ 
\ion{Fe}{xii} & 193.51/195.12 & 0.67 &    0.68 &    0.69 \\ 
\ion{Fe}{xii} & 192.39/195.12 & 0.31 &    0.33 &    0.33 \\ 
\ion{Fe}{xii} & 186.88/196.64 & 3.49--4.10 &    3.86 &    3.79 \\ 
\ion{Fe}{xii} & (186.88+195.12)/249.39 & 17.7--28.1 &   16.35 &   16.79 \\ 
\ion{Fe}{xiii} & 209.92/202.04 & 0.18 &    0.24 &    0.23 \\ 
\ion{Fe}{xiii} & 201.13/(202.04+203.80+203.83) & 0.17--0.18 &    0.22 &    0.08 \\ 
\ion{Fe}{xiii} & 204.94/201.13 & 0.30 &    0.32 &    0.36 \\ 
\ion{Fe}{xiii} & 201.13/197.43 & 4.08 &    5.96 &    4.50 & $\blacktriangleleft$ \\ 
\ion{Fe}{xiii} & 204.94/197.43 & 1.21 &    1.92 &    1.61 \\ 
\ion{Fe}{xiii} & 209.62/200.02 & 0.68--0.72 &    0.78 &    0.94 & \\ 
\ion{Fe}{xiii} & 246.21/251.95 & 0.52 &    0.53 &    0.57 \\ 
\ion{Fe}{xiii} & 261.74/251.95 & 0.097--0.127 &    0.11 &    0.10 \\ 
\ion{Fe}{xiii} & 261.74/201.13 & 0.155--0.186 &    0.18 &    0.17 \\ 
\ion{Fe}{xiii} & 251.95/201.13 & 1.36--1.74 &    1.64 &    1.73 \\ 
\ion{Fe}{xiii} & 251.95/204.94 & 4.58--5.86 &    5.07 &    4.84 \\ 
\ion{Fe}{xiv} & 252.20/264.79 & 0.23 &    0.19 &    0.15 & \\ 
\ion{Fe}{xiv} & 257.39/270.52 & 0.68 &    0.53 &    0.38 & $\blacktriangleleft$\\ 
\ion{Fe}{xiv} & 289.15/274.20 & 0.089\tablenotemark{b} &   0.113 &   0.076 & $\blacktriangleleft$\\ 
\ion{Fe}{xiv} & 274.20/211.32 & 0.71--0.69 &    0.93 &    0.91 \\ 
\ion{Fe}{xiv}\tablenotemark{a} & 270.52/274.20 & 0.45--0.58 &    0.49 &    0.66  & \\ 
\ion{Fe}{xvi} & 251.06/262.98 & 0.57 &    0.52 &    0.52 \\ 
\ion{Fe}{xvi} & 265.00/262.98 & 0.098 &   0.090 &   0.116 \\ 
\ion{Fe}{xvii} & 204.67/254.89 & 0.93 &    0.81 &    1.01 \\ 
\ion{Fe}{xxiv} & 192.03/255.11 & 1.85 &    1.69 &    1.97 \\ 
\ion{Si}{x} & 253.79/258.37 & 0.20 &    0.22 &    0.25 \\ 
\ion{Si}{x} & 277.26/271.99 & 0.77 &    0.77 &    0.89 \\ 
\ion{Si}{x} & 277.26/261.06 & 0.68--0.71 &    0.71 &    0.80 \\ 
\ion{S}{x} & 259.50/264.23 & 0.68 &    0.70 &    0.48 & $\blacktriangleleft$\\ 
\ion{S}{x} & 257.15/264.23 & 0.35 &    0.42 &    0.29 & $\blacktriangleleft$\\ 
\enddata
\tablenotetext{a}{Ratio not listed by DZWW25.}
\tablenotetext{b}{Theoretical ratio significantly different from DZWW25 (see main text).}
\end{deluxetable}

%--------------------------
\subsection{The 255--275~\AA\ region and S\,X lines}\label{sec:s10}

\begin{figure}[t]
    \centering
    \includegraphics[width=1.0\linewidth]{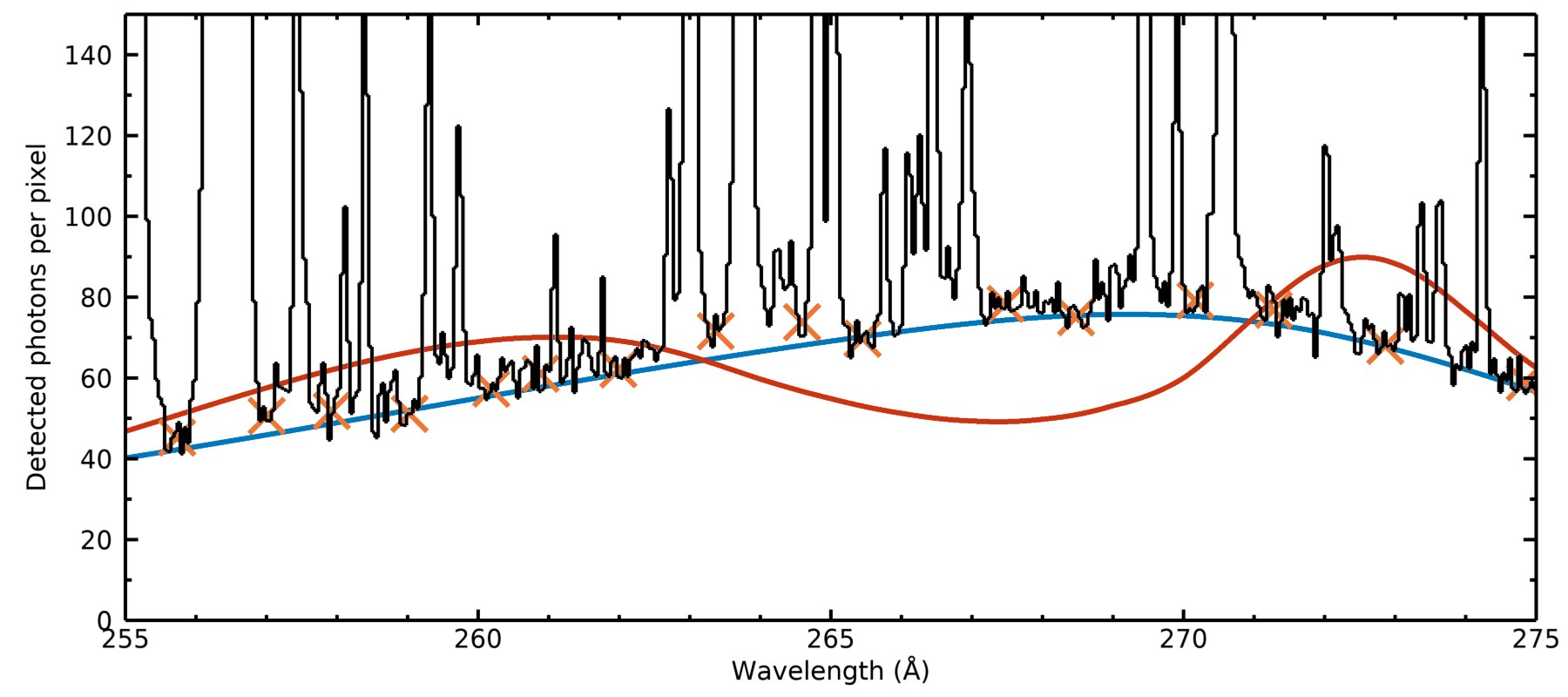}
    \caption{The EIS LW flare spectrum is shown in black, and the measured continuum values are denoted by orange crosses. The blue curve shows the continuum derived from the EIS DEM and the present EA solution. The red curve shows the continuum derived assuming the DZWW25 EA curves.}
    \label{fig:264}
\end{figure}

The largest discrepancies between the present EA curves and those of DZWW25 are found in the 255--275~\AA\ region. Figure~\ref{fig:264} shows the present flare spectrum in this region and the continuum spectrum derived from the present analysis (Section~\ref{sec:cont-calc}). The red line shows the continuum solution if the DZWW25 EA curve is used. It is clear that there is no evidence in the continuum distribution for the EA peaks at 261~\AA\ and 273~\AA.

The DZWW25 EA curves were derived using line ratios, and thus it may be expected that insensitive ratios involving lines formed in the 255--275~\AA\ region would show better agreement using the DZWW25 EA curves compared to the present curves, however this is not the case.
The \ion{Fe}{xiv} 270.52/274.20 and \ion{Fe}{xvi} 265.00/262.98 ratios are located where the DZWW25 EA curve is strongly varying but show better agreement with theory with the present EA curve (Table~\ref{tbl:ratios}). The \ion{Fe}{xiv} 252.20/264.79 and 257.39/270.52 branching ratios also show better agreement with the present EA curve, as do the \ion{Si}{x} 277.26/271.99 and 277.26/261.06 ratios.

The 255--275~\AA\ region contains three \ion{S}{x} lines at 257.15~\AA, 259.50~\AA\ and 264.23~\AA\ that are insensitive to each other, and DZWW25 found that the \ion{S}{x} 264.23~\AA\ line becomes anomalously weak in EIS spectra obtained in 2021 and 2022. This was partly responsible for driving the EA structure seen in the DZWW25 curve in this period. From the 2024 quiet Sun spectrum used here, however, the present EA curve shows good agreement with theory for the two \ion{S}{x} ratios (Table~\ref{tbl:ratios}), whereas discrepancies are found with the DZWW25 curves. 

Figure~17 of DZWW25 demonstrated the time dependence of the \ion{S}{x} problem by considering the \ion{Si}{x} 261.06/\ion{S}{x} 264.23 ratio, which they found to be close to one in quiet Sun conditions for most of the mission but then rose sharply in 2021. We have independently measured this ratio from the datasets 20210616\_211311, 20211013\_031242 and 20220601\_042513 used by DZWW25 and find no evidence for this effect. Considering spectra calibrated with the pre-launch EA curves, DZWW25 found ratios of 1.5, 3.2 and 4.7, whereas we find 1.4, 1.1 and 1.5. G.~Del Zanna (2026, private communication) confirmed that the wrong line had been fit for these datasets.

%===========================
\section{Element Abundances}\label{sec:xsm}

The present analysis has revealed an inverse FIP (I-FIP) effect in the flare region that is required to fit both the low-FIP emission line intensities and the high-FIP continuum intensity. The continuum emission mostly comes from plasma at the peak of the DEM distribution (Figure~\ref{fig:dem}(b)), and the DEM shape in this region is mostly set by the iron ions. Thus the I-FIP effect is effectively a reduction of the Fe/H ratio below photospheric values. 

The I-FIP effect is an uncommon result that goes against the usual solar paradigm of enhanced low-FIP element abundances in the corona \citep{2015LRSP...12....2L}. The first I-FIP measurement for the Sun was presented by \citet{2015ApJ...808L...7D} who studied argon and calcium emission lines, which are classed as high-FIP and low-FIP, respectively. The I-FIP effect was found near sunspots during two flares, and  the most extreme value they found was a Ca/Ar FIP bias of 0.14. At X-ray wavelengths it is possible to obtain abundance measurements of emission lines relative to the continuum in flares, and the first instance of the I-FIP effect was presented by \citet{2020ApJ...891..126K}, who found a Si/H FIP bias of 0.7 in four large solar flares. Other examples of the I-FIP effect from X-ray data were presented in \citet{2024ApJ...972..123N}, who found Fe/H FIP biases of 0.56 and 0.65 for two flares.

As the present result is the first time that the line-to-continuum method has been used to derive absolute abundances from EIS spectra, we have performed a check on the I-FIP result by considering data from the Solar X-ray Monitor \citep[XSM:][]{2020SoPh..295..139M,Mithun_2021ExA....51...33M} on board the Chandrayaan-2 spacecraft. XSM is a compact solar X-ray spectrometer of the type that has been flown on a number of missions over the past 30 years \citep[e.g.,][]{2021FrASS...8...50Y}. The energy resolution is sufficient to enable element abundance measurements through global fitting of the spectrum (both continuum and emission lines) in flares, as demonstrated by \citet{2015ApJ...803...67D} and \citet{2018SoPh..293...21M}. These instruments have no spatial resolution, and so the spectrum is an average over the flare region, but abundances can be obtained at a high cadence of around 1~min or better \citep{Mithun_2022ApJ...939..112M}.

XSM data have been used to derive element abundances in several papers \citep{2021ApJ...912L..12V,2021ApJ...920....4M, 2022ApJ...934..159D,Mithun_2022ApJ...939..112M,Mondal_2023ApJ...955..146M,Nama_2023SoPh..298...55N,yamini_2023ApJ...958..190R,2025arXiv251002102M} for a wide range of solar conditions, including quiet Sun, active regions, microflares and flares. Of these articles, only the \citet{2022ApJ...934..159D} analysis of an active region core considered EIS and XSM data together. Although the EIS analysis did not yield abundance estimates, the XSM silicon FIP bias of 2.0--2.4 was found to be in good agreement with an earlier active region core FIP bias measurement from EIS \citep{2013A&A...558A..73D}.

\begin{figure}
    \centering
    \includegraphics[width=1.\linewidth]{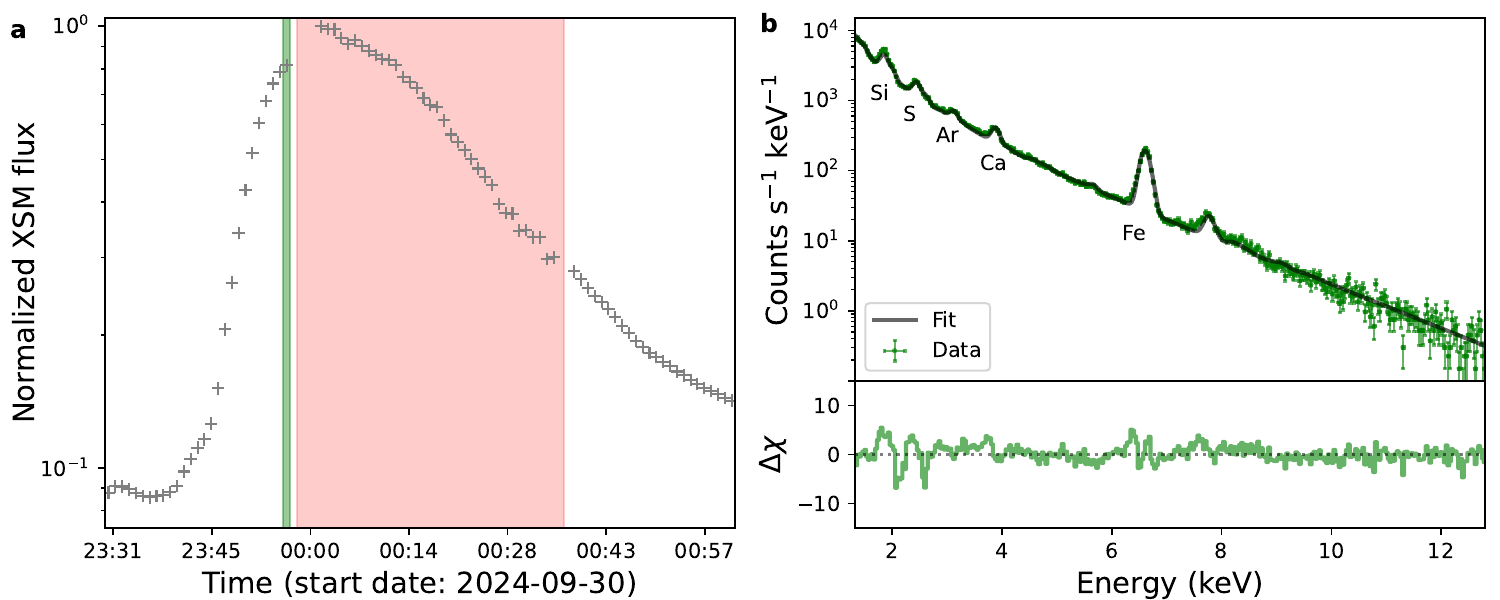}
    \caption{Panel (a) shows the XSM light curve of the flare. The green shaded region indicates the interval 23:56--23:57 UT, and the red shaded region indicates the measurements taken with the attenuator. Panel (b) shows the fitted spectrum for 23:56--23:57 UT and the bottom panel shows the residual between data and model.}
    \label{fig:xsm-lc}
\end{figure}

Figure~\ref{fig:xsm-lc}(a) shows the XSM light curve of the September 30 flare, summed over the instrument's spectral range. The red shaded part of the light curve highlights the period when an attenuator was inserted to reduce the photon flux, and the green line corresponds to the 1~min interval when the EIS flare exposure  was obtained. 
The XSM spectrum was integrated over the latter time period by processing the level 1 data with the XSMDAS, and the resulting spectrum is shown in Figure~\ref{fig:xsm-lc}(b). A fit to the spectrum was performed by creating a spectral model with CHIANTI that consists of two isothermal components plus an additional component for the fluorescent Fe K$\alpha$. The latter is needed as the coronal X-ray spectrum irradiates the photosphere leading to inner-shell excitation of low-ionization iron. When the inner shell electron is filled for these ions, emission at 6.4~keV is produced close to the 6.7~keV feature of \ion{Fe}{xxv}, thus giving a weak bump next to the coronal line \citep{Sarwade_2025}. This is modeled as a Gaussian whose centroid energy and normalization are treated as free parameters during the fitting, while the width is fixed to the XSM energy resolution. Accurately fitting this feature is important for obtaining the iron abundance.

The temperatures and volume emission measures, $\Phi_V$, of the two isothermal components are free to vary, along with the abundances of Si, S, Ca and Fe, which are assumed to be the same for both components. The values for these parameters obtained from the spectrum fit are given in Table~\ref{tbl:xsm}. The abundance, $\epsilon$, is given  relative to hydrogen on a logarithmic scale where hydrogen has an abundance of 12. The FIP bias, $f$,  is the ratio of the derived abundances to the photospheric abundances of \citet{2021A&A...653A.141A}. The iron FIP bias is very close to that found from the EIS analysis. 

The EIS flare exposure used for the continuum analysis was close in time to the flare peak (Figure~\ref{fig:xsm-lc}(a)), when the loop seen in the AIA 131~\AA\ image (Figure~\ref{fig:context}(a)) was very bright. This loop would be expected to be the dominant contributor to the XSM emission.
Based on the derived temperatures and emission measures, the XSM iron abundance is mostly driven by \ion{Fe}{xxiv} and \ion{Fe}{xxv}, which dominate the emission at 6.7~keV. The loop top is significantly brighter in the EIS and AIA images (Figure~\ref{fig:context}) compared to the loop leg used for the EIS analysis, and would be expected to be hotter based on simple 1D loop models \citep[e.g.,][]{1978ApJ...220..643R}. Since XSM sees the entire loop then the spectrum is likely dominated by the loop top, which would explain the high temperature component at $\log\,(T/\mathrm{K})=7.3$ that is not apparent in the EIS DEM distribution (Figure~\ref{fig:dem}).
Thus the XSM iron measurement likely corresponds to the loop top rather than the loop leg measured by EIS. The XSM and EIS results thus suggest the entire loop may exhibit the I-FIP effect in contrast to the result of \citet{2015ApJ...808L...7D} who found the effect was highly localized in a flare.

\begin{deluxetable}{lcc}[t]
\tabletypesize{\normalsize}
\tablecaption{XSM spectrum fitting results. \label{tbl:xsm}}
\tablehead{
  \colhead{Quantity} & 
  \colhead{Value}
} 
\startdata
$\log\,(T_1/\mathrm{K})$  & $6.92^{+0.008}_{-0.006}$ \\
$\log\,(\Phi_{\mathrm{V,}1}/\mathrm{cm}^{-3})$ & $49.49^{+0.007}_{-0.009}$ \\
$\log\,(T_2/\mathrm{K})$  & $7.33^{+0.002}_{-0.002}$ \\
$\log\,(\Phi_{\mathrm{V,}2}/\mathrm{cm}^{-3})$ & $49.32^{+0.005}_{-0.005}$ \\
$\epsilon(\mathrm{Si})$ & $7.37^{+0.02}_{-0.02}$  \\
$f(\mathrm{Si})$ & 0.73 \\
$\epsilon(\mathrm{S})$ & $6.97^{+0.01}_{-0.01}$  \\
$f(\mathrm{S})$ & 0.71 \\
$\epsilon(\mathrm{Ca})$ & $6.59^{+0.02}_{-0.03}$  \\
$f(\mathrm{Ca})$ & 1.95 \\
$\epsilon(\mathrm{Fe})$ & $7.20^{+0.02}_{-0.02}$  \\
$f(\mathrm{Fe})$ & 0.55 \\
\enddata
\end{deluxetable}

%===================
\section{Discussion}\label{sec:discussion}

The present work gives updated EIS effective area (EA) curves for a single date (2024 September 30). 
To derive time-dependent EA curves across the EIS mission lifetime with the continuum will require a number of long-exposure spectral atlas flare observations spaced in time. Three  datasets that may be suitable  were obtained on 2012 March 9 at 03:09~UT  \citep{2013ApJ...767...55D}, 2015 June 21 at 02:08~UT and 2023 May 10 at 14:07~UT \citep[both studied by][]{2025MNRAS.544.2513D}. A follow-on project to the present work will be to attempt the same analysis on these datasets and assess how the effective area curves have changed with time. In the meantime it is recommended that the current effective area curves are used for datasets obtained since 2022 April (the end-point of the DZWW25 analysis period).

DZWW25 highlighted the trade-off between wavy and smooth EA curves. The former come about through having a large number of spline node points that are effective in yielding emission line ratios that match the theoretical predictions from CHIANTI. Enforcing smoothness of the EA curves removes some of the structure but also yields worse comparisons with theory. The comparison in Figure~\ref{fig:ea-comp} suggests that the DZWW25 curves are too far towards the wavy solutions. A solution could be to reprocess the authors' data to enforce greater smoothness to make them more consistent with the continuum solution.

%================
\section{Summary}\label{sec:summary}

New effective area (EA) curves for the EIS SW and LW channels have been calculated from the EUV continuum emission measured from a flare observed on 2024 September 30. In comparison to the curves recently presented by DZWW25, the relative calibration between the SW and LW channels is consistent. However, 
the present EA curves do not show the double-peak structure for the LW channel found by DZWW25, nor the fine-scale structure for the SW channel. The DZWW25 double-peak structure found in the LW 255--275~\AA\ region is found to give worse agreement for insensitive line ratios than the present curve for datasets in 2024, suggesting the EA curves have changed since the DZWW25 analysis period. Consequently, it is recommended that the present EA curves are used for the analysis of datasets obtained since 2022 April 1 (the end of the DZWW25 analysis period). Based on the insensitive line ratio results, caution should be used in using the new EA curves in the wavelength regions below 180~\AA\ and above 285~\AA, where the EA values are very low.

The EIS analysis also yielded an element abundance FIP bias of 0.57, which arises from the spectral model matching both the continuum and emission line intensities.  Since the peak of the derived DEM distribution is mostly driven by iron, then we interpret this as an Fe/H FIP bias that applies at the DEM peak temperature of 10~MK. Analysis of XSM spectra obtained at the same time as the EIS observation yielded an Fe/H ratio of 0.55, giving strong support to the analysis method used for the EIS data. In addition, since the XSM spectrum likely mostly comes from the bright loop top rather than the loop leg used for the EIS analysis, this suggests that the entire flare loop may show the inverse-FIP effect in iron. This work has demonstrated that combined EUV and X-ray modeling of emission lines and continuum is a powerful one for studying the FIP effect in solar flares.

\begin{acknowledgements}
   We thank the anonymous referee for a number of comments and suggestions that have improved the article. We also thank Micah Weberg, Enrico Landi, Harry Warren and Giulio Del Zanna for comments on an early draft of the  manuscript. The author acknowledges funding from the GSFC Internal  Scientist Funding Model competitive work package program, and the \hinode\ project. \hinode\ is a Japanese mission developed and launched by ISAS/JAXA, with NAOJ as domestic partner and NASA and STFC (UK) as international partners. It is operated by these agencies in co-operation with ESA and NSC (Norway).
\end{acknowledgements}

\facilities{Hinode(EIS), SDO(AIA)}, Chandrayaan-2(XSM)

\bibliography{ms}{}
\bibliographystyle{aasjournal}

\appendix

%====================================================
\section{Contribution of weak lines to the continuum}\label{app:pseudo}

The CHIANTI spectral models yield over 100,000 emission lines in the wavelength ranges of both the EIS SW and LW channels. Many of these lines have negligible intensity, but together they are expected to contribute a pseudo continuum across the EIS wavelength bands. This means that the continuum measured from the EIS spectrum (Figure~\ref{fig:cont-fit}) will be an over-estimate of the true continuum. In this section we assess  the magnitude of this effect.

Synthetic spectra are generated with CHIANTI using the DEM derived from the emission line analysis (Figure~\ref{fig:dem}). For the call to the CHIANTI routine \textsf{make\_chianti\_spec}, the keyword \textsf{/all} is set, which includes every CHIANTI transition in the final spectrum. Many weak atomic transitions in CHIANTI do not have experimental wavelengths and so the wavelengths are obtained from theoretical calculations. The accuracy of these wavelengths varies considerably: Some may be accurate to a fraction of an Angstrom, while others may be uncertain to several Angstroms. For estimating the pseudo continuum, it is important to include all of these weak transitions.

\begin{figure}[t]
    \centering
    \includegraphics[width=\linewidth]{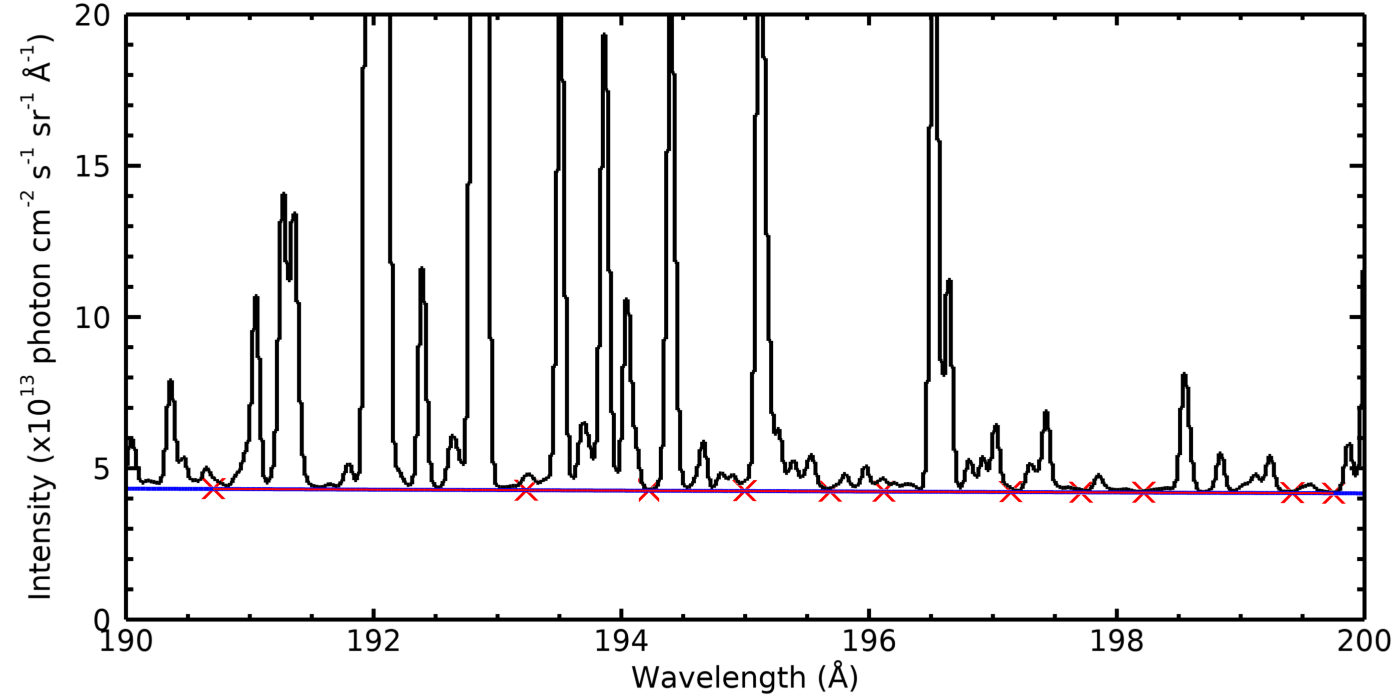}
    \caption{A CHIANTI synthetic spectrum derived from the flare DEM (Figure~\ref{fig:dem}). The blue line shows the continuum component of the spectrum, and the red crosses show the wavelengths at which the flare continuum intensity was estimated from the EIS spectra (Figure~\ref{fig:cont-fit}).}
    \label{fig:pseudo}
\end{figure}

Figure~\ref{fig:pseudo} shows the CHIANTI synthetic spectrum and the true continuum is indicated by the blue line. The red crosses show the wavelengths at which the continuum intensity was estimated from the EIS spectrum, as described in Section~\ref{sec:cont-meas} and shown in Figure~\ref{fig:cont-fit}. At most locations the emission line component to the spectrum is small compared to the continuum. To quantify this, the synthetic spectrum is averaged over the same wavelength windows used for the EIS continuum analysis and compared with the true continuum averaged over these windows. For the SW channel, the median enhancement of the spectrum over the true continuum is 5.6\%\ for 40 wavelength locations. The largest enhancement is 113\%\ at 203.5\,\AA. This is because CHIANTI predicts an \ion{Fe}{xviii} line at this location that is not present in the EIS spectrum -- see discussion in \citet{2025MNRAS.544.2513D}. Nine other locations have enhancements between 10\%\ and 18\%. For the LW channel, the median enhancement is 4.4\%\ for 41 wavelength locations. The largest enhancement is 27\%\ at 257.9~\AA, and four other locations have enhancements between 10\%\ and 14\%. There is no significant wavelength trend to the enhancements across the SW or LW channels.

CHIANTI does not contain all of the emission lines that contribute to the EIS wavebands, but it is mostly complete for transitions involving atomic states with principal quantum numbers ($n$) 2 and 3 that dominate in this part of the spectrum. To demonstrate that transitions with $n\ge 4$ are not significant, the above calculations were repeated excluding all transitions involving $n=4$ states. Of the 81 wavelength locations, omitting the $n=4$ transitions reduces the pseudo continuum  by $<$~1\,\%\ in 71 locations, and by $>10$\,\%\ in two locations. The median enhancements over the true SW and LW continua caused by the pseudo continua are unchanged when omitting the $n=4$ transitions. Most of the transitions missing from the CHIANTI models are expected to be from states with $n\ge 5$, that will be weaker than the transitions from $n=4$ states. We therefore conclude that, given our current knowledge of the EUV atomic data, the missing transitions will not significantly impact the pseudo continuum estimated here. 

The above analysis suggests that the measured EIS continuum values over-estimate the true continuum values by around 5\%. As there is no wavelength trend to the enhancements, then the pseudo continuum does not impact the derived effective area curves (Figure~\ref{fig:ea-comp}) but it would impact the FIP bias value of 0.57 obtained in Section~\ref{sect:dem}, reducing it by 5\%.

%-------------------------------------------------
\section{Background subtraction for the continuum}\label{app:bg}

A region to the north of the flare was chosen to create a background spectrum that was subtracted from the flare spectrum (Section~\ref{sec:prep} and Figure~\ref{fig:cont-bg}) to yield the true flare continuum. The background spectrum accounts for both the true continuum from the background corona as well as the detector background. Here we demonstrate that the detector background is independent of the location on the detector, and hence it is appropriate to use a measurement at a different $y$ position to estimate the background at the flare location.

The detector background consists of the CCD pedestal, the dark current, read noise, and any residual warm pixels that are not removed by \textsf{eis\_prep}. It can be estimated from dark frames (``darks"): exposures taken with the shutter closed such that no solar photons are detected. Darks are taken on a weekly basis with EIS, and one was obtained on 2023 October 1 around 17~UT, less than 24 hours after the flare observation. The procedure is to run the engineering study \textsf{REGCAL081} to obtain exposures from the bottom halves of the EIS CCDs followed by \textsf{REGCAL082} for the top halves.  Since the flare dataset was obtained entirely from the top half of the CCD, only the latter was used.

\begin{figure}[h]
    \centering
    \includegraphics[width=1.0\linewidth]{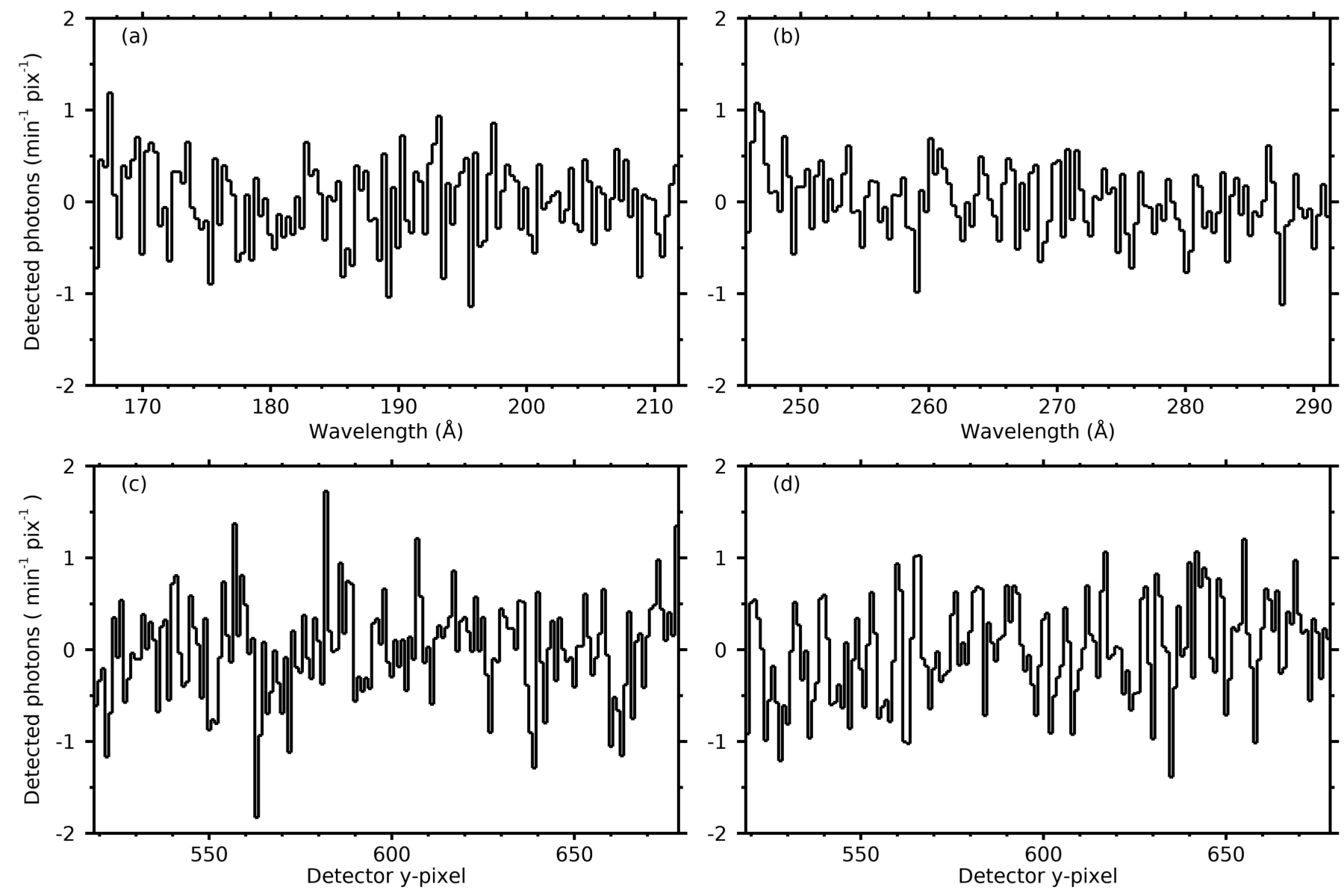}
    \caption{Variations of the average dark current in the wavelength (Panels (a) and (b)) and detector $y$ (Panels (c) and (d)) directions, for the EIS SW and LW channels. The data are from a dark frame obtained on 2024 October 1.}
    \label{fig:dark}
\end{figure}

\textsf{REGCAL082} was run between 17:10~UT and 17:16~UT, producing two exposures of duration 150~s. The data were processed with \textsf{eis\_prep} into photon units, and results from the first exposure are shown here. Figure~\ref{fig:dark} shows the variation of the detector background with wavelength (Panels (a) and (b)) and detector $y$ location (Panels (c) and (d)). For Panels (a) and (b), the data were averaged over the 160 $y$ pixels that match those observed with the flare dataset, and binned by 16 pixels in the wavelength direction. The intensity was normalized such that the average intensity is zero, and the units are given as the photons obtained in 60~s per unbinned pixel. The plots  can thus be directly compared to the flare spectra shown in  Figure~\ref{fig:cont}.
For Panels (c) and (d), the intensity was averaged over the entire wavelength range for each channel, and the displayed $y$ range corresponds to the pixel range used for the flare dataset.

The intensity variations seen in Figure~\ref{fig:dark} arise from the intrinsic noise of the CCDs (dark current and read noise) and the randomly-distributed residual warm pixels. It can be seen that the trends in  the wavelength and $y$ directions are less than 1 photon from end-to-end,  hence it is reasonable to assume the detector background is constant over the detector.

\end{document}